\documentclass{article}
\usepackage{preamble}
\usepackage{arxiv}
\title{Addressing Confounding and Continuous Exposure Measurement Error Using Corrected Score Functions}

\author{Brian D. Richardson$^{1,\dag}$, Bryan S. Blette$^{2,\dag}$, Peter B. Gilbert$^{3}$, and Michael G. Hudgens$^{1,*}$ \\ 
$^{1}$Department of Biostatistics, University of North Carolina at Chapel Hill
\\ Chapel Hill, NC, U.S.A. \\
$^{2}$Department of Biostatistics, Vanderbilt University Medical Center
\\ Nashville, TN, U.S.A. \\
$^{3}$Department of Biostatistics, University of Washington and Fred Hutchinson Cancer Center
\\ Seattle, Washington, U.S.A. \\
$^\dag$Indicates co-first authors \\
$^*$Email: mhudgens@email.unc.edu}

\begin{document}

\maketitle

\begin{abstract}
Confounding and exposure measurement error can introduce bias when drawing inference about the marginal effect of an exposure on an outcome of interest. While there are broad methodologies for addressing each source of bias individually, confounding and exposure measurement error frequently co-occur, and there is a need for methods that address them simultaneously. In this paper, corrected score methods are derived under classical additive measurement error to draw inference about marginal exposure effects using only measured variables. Three estimators are proposed based on g-formula, inverse probability weighting, and doubly-robust estimation techniques. The estimators are shown to be consistent and asymptotically normal, and the doubly-robust estimator is shown to exhibit its namesake property. The methods, which are implemented in the R package \texttt{mismex}, perform well in finite samples under both confounding and measurement error as demonstrated by simulation studies. The proposed doubly-robust estimator is applied to study the effects of two biomarkers on HIV-1 infection using data from the HVTN 505 preventative vaccine trial.
\end{abstract}


\newpage

\section{Introduction}
\label{s:intro}

Consider estimating the effect of a continuous exposure on an outcome when (i) the exposure is measured with error and (ii) the exposure-outcome association is potentially confounded. A motivating example is provided by the HVTN 505 trial of a preventive HIV vaccine. This HIV vaccine efficacy trial stopped administering immunizations early after reaching predetermined cutoffs for efficacy futility~\citep{hammer2013}. However, subsequent analyses of trial data identified several immunologic biomarker correlates of HIV acquisition among HIV vaccine recipients~\citep{janes2017,fong2018,neidich2019}. Some of these biomarkers could be possible target immune responses for future vaccines, so it is of interest to assess the effect of these biomarkers on risk of HIV acquisition. These biomarker effects must be assessed carefully since measurement of the biomarkers is subject to error and the association between the biomarkers and HIV risk is likely confounded.

Methods for estimating continuous exposure effects in the presence of confounding and exposure measurement error are limited. Existing measurement error methods can be used to target \textit{conditional} treatment effects (i.e., the effects of a treatment on individuals conditional on their covariate values) by adjusting for baseline covariates in outcome models, for example, when performing regression calibration~\citep{carroll2006}. However, \textit{marginal} effects (i.e., the average effects of a treatment across all individuals in a population) are often of primary interest. \cite{josey_estimating_2023} introduced a multiple imputation method that combines regression calibration and Bayesian approaches to estimate marginal causal effects in the presence of continuous exposure measurement error. The method in \cite{josey_estimating_2023} requires replicate error-prone measurements from a cluster to estimate potential outcomes at cluster-level exposure values; moreover, it relies on a correctly specified outcome model. This paper develops methods which can be used when a continuous exposure is subject to measurement error and neither replicate measurements nor validation data are available. The approach is based on the corrected score (CS) function method for estimation and inference in the presence of measurement error~\citep{carroll2006}. Three CS methods targeting marginal causal estimands are proposed, based on the g-formula, inverse probability weighting (IPW), and doubly-robust (DR) estimation.

This paper proceeds as follows. In Section~\ref{s:notation} notation and the target estimand are defined, and assumptions are stated. In Section~\ref{s:methods} the three proposed estimators are introduced. In Section~\ref{s:simulation} the proposed estimators are evaluated in a simulation study, and in Section~\ref{s:application} one of the estimators is applied to study two biomarkers from the HVTN 505 vaccine trial. Section~\ref{s:discussion} concludes with a discussion of the advantages and limitations of the proposed methods.

\section{Notation and Estimand}
\label{s:notation}

Suppose there are $m$ exposures/treatments of interest which may or may not be measured with error. Let $\bA = (A_{1}, A_{2}, ..., A_{m})$ be a row vector denoting the true exposure values for an individual and $\bAs = (\As_{1}, \As_{2}, ..., \As_{m})$ be the corresponding measured exposure values. For example, in the HVTN 505 trial, one biomarker of interest $A$ is antibody-dependent cellular phagocytosis activity. This biomarker was not observed exactly, but an imperfect phagocytic score $\As$ was measured. Assume a classical additive measurement error model, i.e., that $\bAs = \bA + \beps_{me}$, where the measurement error $\beps_{me} = (\epsilon_{me_1}, \dots, \epsilon_{me_m})$ follows a multivariate normal distribution $\bsN(\bo, \bSigma_{me})$ with known covariance matrix $\bSigma_{me}$. If some components of $\bAs$ are perfectly measured, then the corresponding elements of $\bSigma_{me}$ are zero. The exposures may be either discrete or continuous, but all mismeasured exposures are assumed to be continuous.

Let $Y$ be the outcome of interest. Define $Y(\ba)$ to be the potential outcome under exposure $\bA = \ba = (a_{1}, a_{2}, ..., a_{m})$. Assuming there is at least one continuous exposure, each individual has an infinite number of potential outcomes. Let $\bL =  (L_{1}, L_{2}, ..., L_{p})$ represent a vector of baseline covariates measured prior to exposure. This covariate set is assumed to be sufficient to adjust for confounding in the sense that the conditional exchangeability assumption described below is satisfied. Assume that $n$ i.i.d. copies of the random variables $(\bL, \bAs, Y)$ are observed.

The estimand of interest is the mean dose-response surface, namely $\eta(\ba) \equiv \E\{ Y(\ba) \}$ for $\ba \in \sA$, where $\sA$ represents the $m$-dimensional space of exposure values of interest. For example, with one exposure, $\eta(a)$ may be a dose response curve across an interval of exposure values. Each of the proposed estimators described in this paper will make assumptions that explicitly or implicitly impose restrictions on this surface. The proposed IPW estimator will directly target the parameters of a marginal structural model (MSM) given by $\eta(\ba) = \eta(\ba; \bgam)$ where $\bgam$ is a row vector quantifying the causal effects of the exposures on the outcome. For example, the MSM could be assumed to have the form $\eta(\ba; \bgam) = g^{-1}(\gamma_0 + \bgam_a \ba^{T})$ for $\bgam = (\gamma_0, \bgam_a)$ and a monotone link function $g$ (e.g., the logit function). The MSM can also include nonlinear functions of the exposure, e.g., $\eta(a; \bgam) = g^{-1}(\gamma_0 + \gamma_1 a + \gamma_2 a^2)$. In this case, the proposed CS methodology only needs to account directly for measurement error in $A$, not in $A^2$ or other functions of $A$ included in the MSM.

The g-formula and doubly robust estimators assume a parametric model for the mean outcome $\mu(\bL, \bA) \equiv \E(Y | \bL, \bA)$ given the exposure and covariates: $\mu(\bL, \bA) = \mu(\bL, \bA; \bbeta)$, where $\bbeta$ is  row vector of parameters quantifying the association between $(\bL, \bA)$ and $Y$. For example, this outcome model could be assumed to have the form $\mu(\bL, \bA; \bbeta) = g^{-1}(\beta_{0} + \bm{L}\bbeta_{l}^T + \bA\bbeta_{a}^T + \bm{A}\bbeta_{al}\bL^{T})$, where $g$ is a monotone link function, $\bbeta_{al}$ is an $m \times p$ matrix of interaction parameters with appropriate elements constrained to zero to include only relevant interactions, and $\bbeta = [\beta_0, \bbeta_l, \bbeta_a, \vc(\bbeta_{al})]$. Like the MSM, this outcome model is not limited to linear functions of $\ba$. The assumed mean outcome model implies the marginal structural model $\eta(\ba) = \E\{ \mu(\bL, \ba; \bbeta) \}$. The choice of specifying an MSM either explicitly with $\eta(\ba; \bgam)$ or implicitly with $\mu(\bL, \bA; \bbeta)$ is a key consideration when choosing between methods.

The proposed methods in Section~\ref{s:methods} rely on a standard set of assumptions used in causal inference: (i) causal consistency, $Y = Y(\ba)$ when $\bA = \ba$; (ii) conditional exchangeability, $Y(\ba) \perp \!\!\! \perp \bA | \bL$ for all $\ba \in \sA$; and (iii) positivity, $f_{\bA|\bL}(\ba | \bl) > 0$ for all $\bl$ such that $f_{\bL}(\bl) > 0$ and for all $\ba \in \sA$. Here and below, the notation $f_{X}(x)$ denotes the marginal probability density function (pdf) of a random variable $X$, and $f_{X|W}(x|w)$ denotes the conditional pdf of $X$ given the random variable $W$. In addition, assume that the outcome and covariates are not measured with error, and that there is no model mis-specification unless otherwise stated.

\begin{figure}
\centering
\fbox{\includegraphics[width=\textwidth, trim = 4cm 19.5cm 3cm 2.9cm]{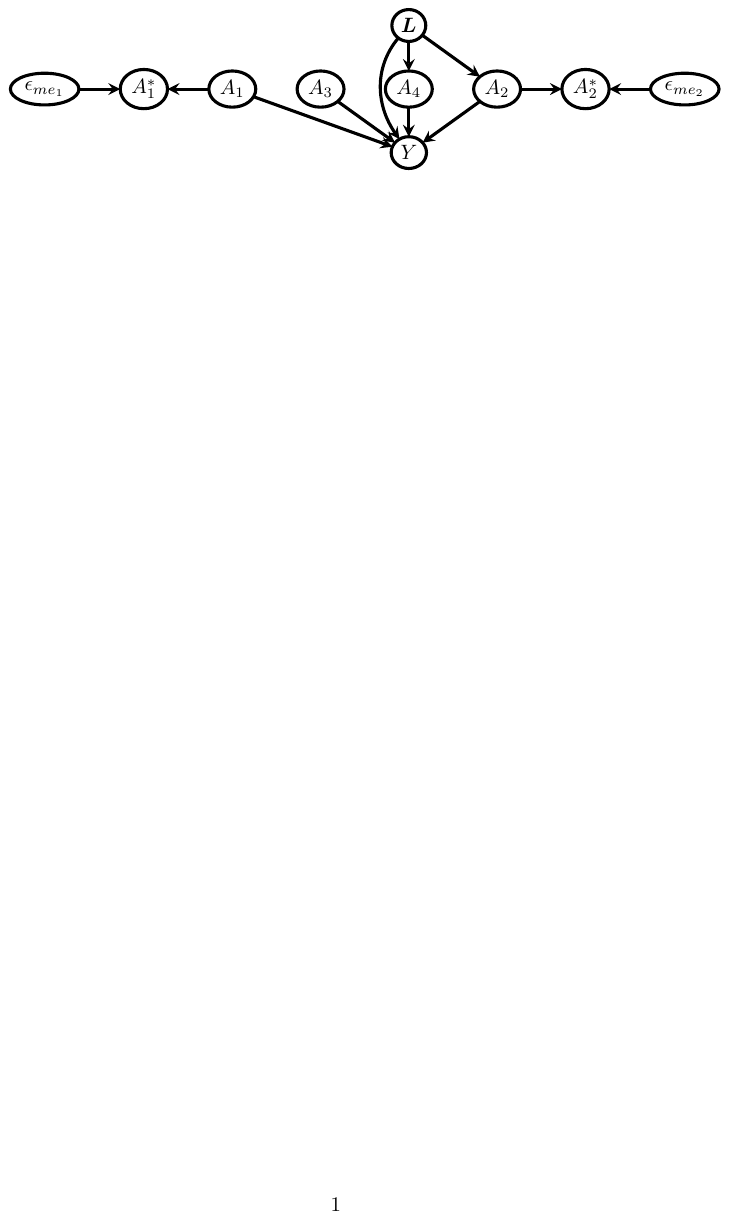}}
\caption{An example directed acyclic graph (DAG), with variables defined in Section~\ref{s:notation} and $\epsilon_{me_{1}}$ and $\epsilon_{me_{2}}$ corresponding to measurement error. This DAG represents a scenario with $m=4$ exposures, one each of the following: mismeasured and unconfounded ($A_{1}$), mismeasured and confounded ($A_{2}$), correctly measured and unconfounded ($A_{3}$), and correctly measured and confounded ($A_{4}$). In this case, $A_{3} = \As_{3}$ and $A_{4} = \As_{4}$ since they are both measured without error.}
\label{fig:one}
\end{figure}

The estimators proposed in this paper are applicable in settings such as the example directed acyclic graph (DAG) in Figure 1. The methods can accommodate all four types of exposures in the DAG: unconfounded and correctly measured, unconfounded and mismeasured, confounded and correctly measured, and confounded and mismeasured. Here and throughout, exposure measurement error is assumed to be independent of $Y, \bA, \bL$. For example, in Figure~\ref{fig:one}, $\beps_{me} \perp \!\!\! \perp Y, \bA, \bL$ since $\As_{1}$ and $\As_{2}$ are colliders on all paths between $\epsilon_{me_{1}}, \epsilon_{me_{2}}$ and $Y, \bA, \bL$.

\section{Methods}
\label{s:methods}

The proposed estimators combine existing methods to (i) adjust for confounding using g-formula, inverse probability weighting, and doubly-robust techniques and (ii) correct exposure measurement error using CS functions. To begin, the CS method is briefly reviewed.

\subsection{Review of Corrected Score Functions}
\label{sec:cs}

Consider an M-estimator $\bthetahat_0$ of a parameter $\btheta$ in the absence of measurement error. That is, $\bthetahat_0$ solves the estimating equation $\sum_{i=1}^n \bPsi_0(Y_i, \bL_i, \bA_i ; \btheta) = \bo$, for some score (or estimating) function $\bPsi_0$ that is unbiased, i.e., $\E\{\bPsi_0(Y, \bL, \bA ; \btheta)\} = \bo$. Note that the traditional parametric IPW, g-formula, and doubly-robust estimators (ignoring measurement error) can all be expressed as M-estimators. Given that the true exposure values $\bA$ are not observed, solving this equation is generally not possible. However, the observed data can in some cases be used to construct a corrected score function $\bPsi_{CS}(Y, \bL, \bAs, \btheta)$ with the property that
\begin{align}
\label{eq:csf_condition}
    \E\left\{\bPsi_{CS}\left(Y, \bL, \bAs ; \btheta \right) | Y, \bL, \bA\right\} =
    \bPsi_0\left(Y, \bL, \bA ; \btheta\right).
\end{align}
By the law of iterated expectations, $\bPsi_{CS}$ is then an unbiased estimating function, and therefore the root $\bthetahat$ of $\sum_{i=1}^{n}\bPsi_{CS}(Y_i, \bL_i, \bAs_i ; \btheta)$ is an M-estimator for $\btheta$.

\cite{novick_corrected_2002} provide one strategy to construct corrected score functions using Monte-Carlo simulation of complex variables, which can be applied to any estimating function $\bPsi_0$ that is conditionally unbiased, i.e., $\E\{\bPsi_0(Y, \bL, \bA ; \btheta) | \bA\} = \bo$. The method entails adding imaginary (i.e., complex-valued) measurement error to the observed $\bAs$, then taking a conditional expectation, given the observed data, of the real component of the complex-valued estimating function. The result is a function $\bPsi_{CS}$ of the observed data that satisfies \eqref{eq:csf_condition}. More formally, let
\begin{align}
\label{eq:csf}
    \bPsi_{CS}\left(Y, \bL, \bAs ; \btheta\right) = \E\left[ \real \left\{ \bPsi_{0}(Y, \bL, \bAt ; \btheta) \right\} |
    Y, \bL, \bAs \right],
\end{align}
where $\bAt = \bAs + i\bepst$, $i = \sqrt{-1}$, $\real(\cdot)$ denotes the real component of a complex number, and $\bepst \sim \bsN(\bo, \bSigma_{me})$ has the same distribution as the real measurement error $\beps_{me}$ and is independent of all other variables. Note $\bPsi_{CS}$ is implicitly also a function of $\bSigma_{me}$, which is assumed known. In some cases, the expectation in \eqref{eq:csf} has a closed form expression (see Section~\ref{sec:eval-cs}). In all cases, it can be approximated stochastically with i.i.d. simulated copies $\bepst_1, \dots, \bepst_B$ of $\bepst$. The Monte-Carlo corrected score (MCCS) function using $B$ replicates is given by $\bPsi_{MCCS}^B(Y, \bL, \bAs ; \btheta) = B^{-1} \sum_{b=1}^B \real \left\{ \bPsi_{0}(Y, \bL, \bAt_b ; \btheta) \right\}$, where $\bAt_b = \bAs + i\bepst_b$. By the weak law of large numbers (conditional on $Y, \bL, \bAs$), $\bPsi_{MCCS}^B(Y, \bL, \bAs ; \btheta)$ converges in probability to $\bPsi_{CS}(Y, \bL, \bAs ; \btheta)$ as $B \to \infty$, so $\bPsi_{MCCS}^B$ can approximate $\bPsi_{CS}$ for a large number of replicates $B$. Note however that for \textit{any} number of replicates $B \geq 1$, $ \bPsi_{MCCS}^B$ satisfies \eqref{eq:csf_condition}, and therefore the root of $\sum_{i=1}^{n}\bPsi_{MCCS}^B(Y_i, \bL_i, \bAs_i ; \btheta)$ is an M-estimator.

\subsection{CS G-formula Estimator}

The first proposed method applies the CS framework to a traditional g-formula estimator. When there is no measurement error, the g-formula estimator entails fitting the outcome model, then using the resulting parameter estimator $\bbetahat$ to estimate the dose-response curve as $\etahat(\ba) = n^{-1} \sum_{i=1}^{n} \mu(\bL_i, \ba; \bbetahat)$. This can be expressed as an M-estimator with estimating function
\begin{align}
\label{eq:gf}
    \bPsi_{0-GF}(Y, \bm{L}, \bA ; \btheta_{GF}) =
    \begin{bmatrix}
       \left\{ Y - \mu(\bL, \bA; \bbeta) \right\}
       \pd{\bbeta}^{T} \mu(\bL, \bA; \bbeta) \\
       \eta(\ba) - \mu(\bL, \ba; \bbeta)
    \end{bmatrix}^{T},
\end{align}
where $\btheta_{GF} = [\bbeta, \eta(\ba)]$, and $\pd{\bbeta} \mu(\bL, \bA; \bbeta)$ is the partial derivative of $\mu(\bL, \bA; \bbeta)$ with respect to $\bbeta$. To estimate the dose response surface in practice, one can compute $\etahat(\ba)$ for a large grid of points $\ba$ in the space of interest. To accommodate exposure measurement error, the proposed CS g-formula function is $\bPsi_{CS-GF}(Y, \bL, \bAs ; \btheta_{GF}) = $
\begin{align}
\label{eq:gf-cs}
    \begin{bmatrix}
       \E \left( \real \left[ \left\{ Y - \mu(\bL, \bAt; \bbeta) \right\}
       %(1, \bL, \bAt, \bL \otimes \bAt)^{T}
       \pd{\bbeta}^{T} \mu(\bL, \bAt; \bbeta)
       \right] | Y, \bL, \bAs \right) \\
       \eta(\ba) - \mu(\bL, \ba; \bbeta)
    \end{bmatrix}^{T},
\end{align}
where $\bAt$ is defined as in Section~\ref{sec:cs}. Note that no correction is needed for the second row of $\bPsi_{CS-GF}$ since it does not involve $\bAs$.

\subsection{CS IPW Estimator}
\label{sec:ipw}

Another common causal inference technique is to weight an estimator by the inverse probability/density of exposure(s) conditional on a set of covariates $\bL$ that satisfy the conditional exchangeability assumption. IPW estimators can also be expressed as M-estimators and, in the absence of measurement error, the IPW estimating function is
\begin{equation}
\label{eq:ipw}
    \bPsi_{0-IPW}(Y, \bL, \bA ; \btheta_{IPW}) =
    \begin{bmatrix}
       \bPsi_{PS}^{T}(\bL, \bA ; \bxi, \bzeta) \\
       SW(\bL, \bA ; \bxi, \bzeta)
       \left\{ Y - \eta(\bA; \bgam) \right\}
       \pd{\bgam}^{T} \eta(\bA; \bgam)
    \end{bmatrix}^{T},
\end{equation}
where
\begin{equation}
\label{eq:sw}
SW(\bL, \bA ; \bxi, \bzeta) = \frac{f_{\bA}(\bA; \bxi)}{f_{\bA|\bL}(\bA | \bL; \bzeta)}
\end{equation}
are stabilized weights, $\bPsi_{PS}$ is an estimating function for fitting propensity score models (i.e., assumed models for $\bA$ and $\bA|\bL$) with parameters $\bxi$ and $\bzeta$, $\btheta_{IPW} = (\bgam, \bxi, \bzeta)$, and $\bgam$ are the MSM parameters.

Heuristically, weighting works by creating a pseudo-population where confounding is no longer present. For continuous exposures, this is accomplished by weighting by the inverse of the joint density of exposures conditional on confounders. The unconditional joint density in the numerator of $SW$ is used to stabilize the weights, and does not need to be correctly specified in order for the estimator to be consistent. The second equation in \eqref{eq:ipw} corresponds to a weighted estimator of the generalized linear regression of the outcome on only the exposure and is indexed by the MSM parameters of interest $\bgam$ rather than the $\bbeta$ parameters from equations \eqref{eq:gf} and \eqref{eq:gf-cs}. This regression model does not include $\bL$ since weighting eliminates confounding. When the weights $SW$ are known, such as in a conditionally randomized experiment, one does not need to fit models for $\bA$ and $\bA|\bL$, and the estimating equation vector \eqref{eq:ipw} can instead be solved with the first equation $\bPsi_{PS}$ removed and the true $SW$ used in the second equation. In observational studies, the weights are usually not known and must be estimated. Details for fitting propensity score models are in Web Appendix A. Importantly, the consistency of the IPW estimator relies on a consistent estimator of $f(\bA|\bL)$; the DR estimator proposed in Section~\ref{s:dr} is one alternative that relaxes this requirement.

To account for mismeasured exposures, the CS IPW function is \\
$\bPsi_{CS-IPW}(Y, \bL, \bAs ; \btheta_{IPW}) =$
\begin{equation}
\label{eq:ipw-cs}
    \begin{bmatrix}
       \bPsit_{PS}^{T}(\bL, \bAs ; \bxi, \bzeta) \\
       \E \left( \real \left[ SW(\bL, \bAt ; \bxi, \bzeta)
       \left\{ Y - \eta(\bAt; \bgam) \right\}
        \pd{\bgam}^{T} \eta(\bAt; \bgam)
       \right] | Y, \bL, \bAs \right)
    \end{bmatrix}^{T},
\end{equation}
where $\bPsit_{PS}$ is an estimating function to fit the marginal and conditional densities $f_{\bA}(\bA; \bxi)$ and $f_{\bA|\bL}(\bA | \bL; \bzeta)$ for the true exposure $\bA$ using the measured $\bAs$. Note these propensity models have a mismeasured \textit{outcome} and, assuming the measurement error is additive and $\bSigma_{me}$ is known, can be fit using existing methods, e.g., regression calibration or simulation-extrapolation \citep{carroll2006}. 

\subsection{CS Doubly-Robust Estimator}
\label{s:dr}

Both g-formula and IPW methods rely on model specifications that may not be correct in practice. The g-formula provides consistent estimation of potential outcome means only when the outcome model conditional on exposures and confounders is correctly specified. Likewise, IPW estimators are consistent only when the propensity score models (and the MSM) are correctly specified. In contrast, doubly-robust (DR) estimators entail specifying both propensity and outcome models, but remain consistent if one model is mis-specified and the other is not (\citealp*{robins1994}; \citealp{lunceford2004,bang2005}). One such DR method is a weighted regression estimator adapted from \cite{zhang2016} for the additive measurement error setting. In general, this method only applies when the outcome model and the implied MSM are both linear on the scale of the same link function. That is, when $\mu(\bL, \bA; \bbeta) = g^{-1}(\beta_{0} + \bm{L}\bbeta_{l}^T + \bA\bbeta_{a}^T + \bm{A}\bbeta_{al}\bL^{T})$ and $\eta(\ba; \bgam) = g^{-1}(\gamma_0 + \bgam_a \ba^{T})$ for the same $g$. This class of models includes two special cases where (i) the outcome model is linear with the identity link or (ii) the outcome model is linear with the log link and no $\bA \times \bL$ interactions.
%    \item[] \textit{Case A}: a linear outcome model (where $g$ is the identity link), which corresponds to a linear MSM with $\gamma_0 = \beta_0 + \E(\bL)\bbeta_{l}^T$ and $\bgam_a = \bbeta_a + \E(\bL)\bbeta_{al}^T$.
%    \item[] \textit{Case B}: a log-linear outcome model (where $g$ is the $\log$ link) without $\bA \times \bL$ interactions (i.e., $\bbeta_{al} = \bo$), which corresponds to a log-linear MSM with $\gamma_0 = \beta_0 + \log[ \E\{ \exp( \bL\bbeta_{l}^T )\}]$ and $\bgam_a = \bbeta_a$.
%\end{itemize}

The DR estimator is similar to the standard g-formula estimator, but with a weighted outcome regression where the weights are the inverse probability weights given in equation \eqref{eq:sw}. With perfectly measured exposures, the DR estimator has corresponding estimating function $\bPsi_{0-DR}(Y, \bL, \bA, \btheta_{DR}) =$
\begin{equation}
\label{eq:dr}
    \begin{bmatrix}
        \bPsi_{PS}^{T}(\bL, \bA ; \bxi, \bzeta) \\
        SW(\bL, \bA ; \bxi, \bzeta)
          \{ Y - \mu(\bL, \bA; \bbeta) \}
          \pd{\bbeta}^{T} \mu(\bL, \bA; \bbeta) \\
          %(1, \bL, \bA, \bL \otimes \bA)^{T} \\
        \eta(\ba) - \mu(\bL, \ba; \bbeta)
    \end{bmatrix}^{T},
\end{equation}
where $\btheta_{DR} = (\btheta_{GF}, \bxi, \bzeta)$. The corresponding CS DR estimating function is then\\
$\bPsi_{CS-DR}(Y, \bL, \bAs, \btheta_{DR}) =$
\begin{equation}
\label{eq:dr-cs}
    \begin{bmatrix}
        \bPsit_{PS}^{T}(\bL, \bAs ; \bxi, \bzeta) \\
        \E \left( \real \left[ SW(\bL, \bAt ; \bxi, \bzeta)
          \{ Y - \mu(\bL, \bAt; \bbeta) \}
          \pd{\bbeta}^{T} \mu(\bL, \bAt; \bbeta)
          \right] | Y, \bL, \bAs \right) \\
         \eta(\ba) - \mu(\bL, \ba; \bbeta)
    \end{bmatrix}^{T}.
\end{equation}

If the stabilized weights $SW$ are unknown, they are estimated with the root of $\bPsit_{PS}$ as described for the CS IPW estimator; if the weights are known, the estimating equation vector \eqref{eq:dr-cs} can instead be solved with the first element $\bPsit_{PS}$ removed. Like the CS g-formula estimator, the CS DR estimator can be evaluated over a grid of values $\ba$ of interest to estimate a dose-response surface. Importantly, the doubly robust property relies on the MSM implied by the outcome model being correct. This condition is always met when the outcome model is correctly specified, but is not always met for an incorrect outcome model. For this reason, the doubly robust estimator is only recommended in scenarios where the specified outcome model is believed to be compatible with the true MSM.

\subsection{Evaluating Corrected Score Functions}
\label{sec:eval-cs}

The three proposed CS methods involve expectations of the form $\E[ \real \{ \bPsi_{0}(Y, \bL, \bAt ; \btheta) \} | Y, \bL, \bAs ]$. The MCCS method described in Section~\ref{sec:cs} for approximating this expectation is convenient since it can be applied in a variety of modeling scenarios without dealing with complex algebra. This simulation-based approach can however be computationally intensive, and, if this is a concern, there are some scenarios where a closed form expression for the CS function exists. Section 7.4.3 of \cite{carroll2006} lists several such scenarios, including when $\bPsi_{0}$ is a score function for (i) normal linear regression with an identity link, (ii) Poisson regression with a log link, or (iii) gamma regression with a log link. If the outcome model falls into any of these three cases, then the closed form expressions in \cite{carroll2006} can be used for the g-formula estimator, bypassing any Monte-Carlo approximation.

This strategy cannot be employed directly for the CS IPW or DR functions since they involve the standardized weights $SW$. In Web Appendix B, it is shown that in the special case where the models for $\bA$ and $\bA | \bL$ are normal with identity links and where the MSM has a linear link, the CS IPW function has a closed form. Similarly, if the same holds for the $\bA$ and $\bA | \bL$ models and if the outcome model has a linear link, then the CS DR function has a closed form expression.

\subsection{Large-Sample Properties and Variance Estimation}
\label{sec:asymp-props}

In Web Appendix C, it is shown that each of the three proposed estimators is consistent and asymptotically normal by showing that their corresponding estimating functions have expectation $\bo$~\citep{stefanski2002}. In addition, the CS DR estimator is shown to be consistent when only one of the propensity and outcome models is correctly specified.

Since each proposed method is an M-estimator, consistent estimators of their asymptotic variances are given by the empirical sandwich variance technique. Estimating equations corresponding to the estimation of weights should be included in the estimating equation vector for each method when computing the sandwich variance estimator. Wald $100(1-\alpha)\%$ confidence intervals (CIs) for the parameters of interest can then be constructed in the usual fashion. The sandwich variance estimator is consistent for the asymptotic variance but tends to underestimate the variance in finite samples, leading to undercoverage of the CI. To alleviate this, a bias-corrected (BC) sandwich variance estimator from \cite{fay_small-sample_2001} can be used.

\subsection{Handling Unknown Measurement Error Covariance}
\label{s:unknown-var}

Although the proposed methods require no individual-level supplemental data, a priori knowledge of the measurement error covariance matrix is required. Sometimes this matrix will be known from properties of the measurement devices (e.g., some bioassays, certain widely studied survey instruments), or an estimate $\bSigmahat_{me}$ of the measurement error covariance is available from a previous study, e.g., in the case of educational tests \citep{boyd_measuring_2013}. This can be viewed as summary-level supplemental data. Other times, however, $\bSigma_{me}$ may need to be estimated; Web Appendix D presents guidelines for estimating $\bSigma_{me}$. When no replicates are available and there is no prior knowledge of $\bSigma_{me}$, the proposed methods can still be used in conjunction with a sensitivity analysis. In some settings, an upper bound on the exposure measurement error covariance may be assumed; for example, the covariance of a correctly measured exposure (if estimated in a prior study) may be a reasonable upper bound on measurement error covariance for the corresponding mismeasured exposure. Once upper bounds are determined, inference may be repeated using the proposed methods for a range of $\bSigma_{me}$ specifications to assess robustness of point estimates and confidence intervals to the degree of assumed measurement error; this procedure is more straightforward when the matrix $\bSigma_{me}$ is small and diagonal, and becomes difficult to interpret as the number of non-zero parameters grows.

\section{Simulation Study}
\label{s:simulation}

The performance of the proposed methods was evaluated in three simulation studies. The first simulation examined the proposed CS g-formula approach in a scenario where confounding and additive exposure measurement error were present. 1000 data sets of $n \in \{400, 800, 8000\}$ individuals were simulated, each with the following variables: a confounder $L$ simulated as a uniform random variable, an exposure $A$ with $A|L \sim \mathcal{N}(L, 0.25)$, and an outcome $Y$ with $Y|A,L \sim \mathcal{N}(0.25A + 0.5A^2 + -0.5A^3 + L, 0.16)$. The exposure was subject to classical additive measurement error with known variance $\sigma_{me}^2 = 0.05$. This corresponds to an exposure reliability of $\Var(A)/\Var(A^*) = 0.87$, and reliabilities of  $0.80, 0.75$ for $A^2, A^3$, respectively.

Dose-response curves were estimated using five methods: (i) oracle g-formula using the true values $A$ (which are unknown in practice), (ii) naive g-formula, which assumes $A^{*}$ is perfectly measured, (iii) regression calibration g-formula, (iv) SIMEX g-formula, and (v) CS g-formula, the proposed estimator. Details of the regression calibration and SIMEX estimators used in the simulation studies are provided in Web Appendix E.

\begin{figure}
\centering
\includegraphics[width=6.5in]{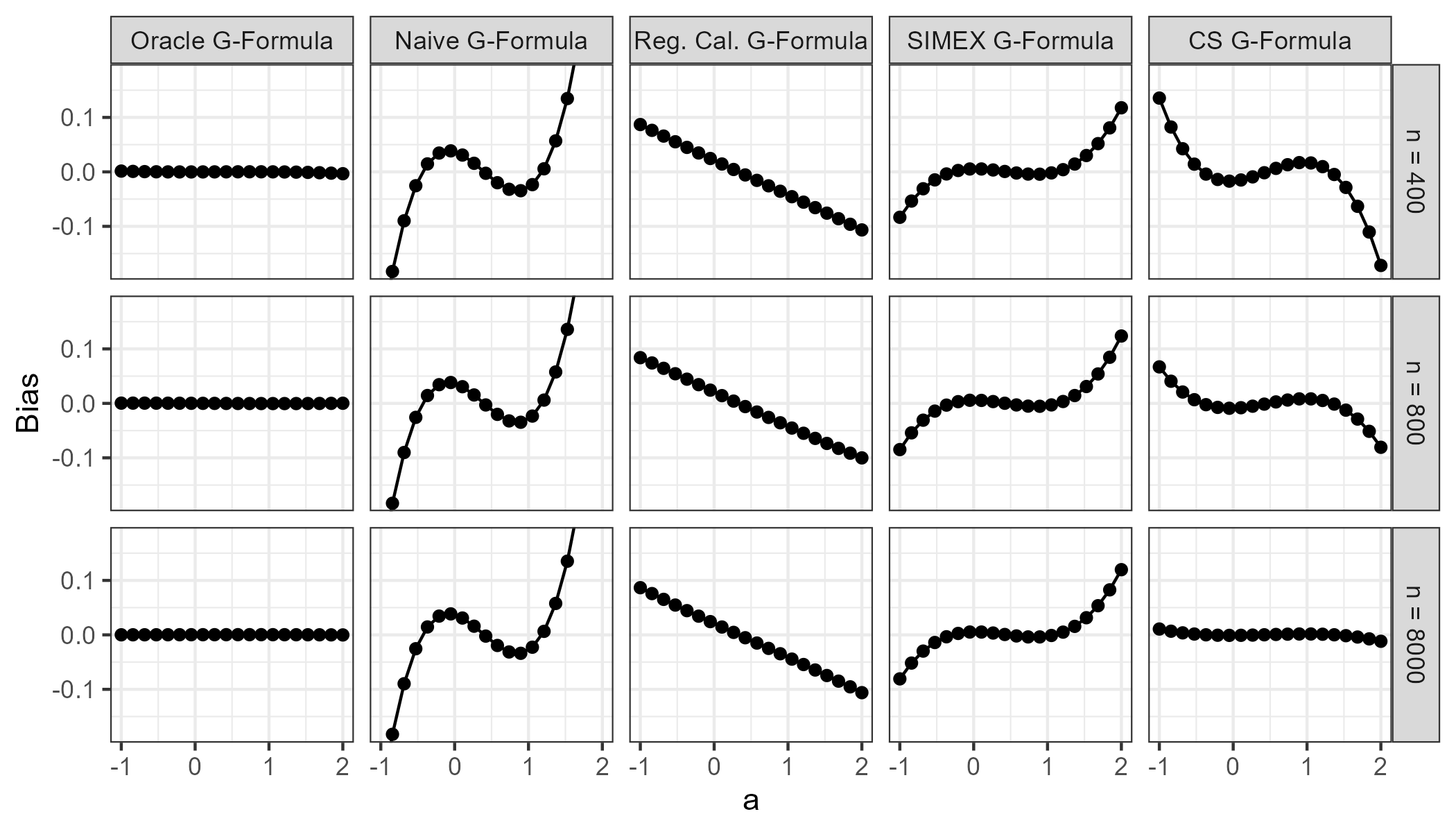}
\caption{Estimated dose-response curve bias for the oracle, naive, regression calibration, SIMEX, and CS g-formula estimators in the first simulation study. Bias refers to the average bias across 1000 simulated data sets for each method evaluated at each point on the horizontal axis, corresponding to setting the true exposure to $a \in [-1, 2]$. For the naive estimator, biases outside of $[-0.18, 0.18]$ are excluded from the plot.}
\label{fig:two}
\end{figure}

To demonstrate performance of the proposed estimator for the entire dose-response curve, empirical biases of the five methods over a grid of points $a \in [-1, 2]$ are displayed in Figure~\ref{fig:two}. The oracle g-formula estimator, which serves as a benchmark since it uses the unobserved $A$, was essentially unbiased. The naive, regression calibration, and SIMEX g-formula method exhibited bias across the range of exposure values considered that did not decrease with increasing sample size. The proposed CS g-formula estimator was approximately unbiased, with some bias towards the boundary of the region of interest which diminished with increasing $n$. Additional simulation results including estimated standard error performance and confidence interval coverage are in Web Appendix F.

The second simulation study examined the proposed CS IPW approach. A total of 1000 data sets of $n \in \{400, 800, 8000\}$ individuals were simulated, each with the following variables: confounder $L \sim \sN(0, 0.36)$, exposure $\bA = (A_1, A_2)$ with $\bA|L$ having multivariate normal distribution with mean $(L^2, -L^2)$ and diagonal covariance matrix with diagonal elements $(1, 1)$, and outcome $Y$ with $Y|L,\bA \sim \sN(A_1 + A_2 + L, 1)$. This implied an MSM of $\eta(\ba;\bgam) = \gamma_{0} + \gamma_{1}a_{1} + \gamma_{2}a_{2}$ for $\bgam = (\gamma_0, \gamma_1, \gamma_2) = (0, 1, 1)$. The exposure was subject to additive measurement error with known diagonal covariance matrix having diagonal elements $(0.2,0.2)$, corresponding to a reliability of 0.84 for each exposure.

\begin{figure}
\centering
\includegraphics[width=6.5in]{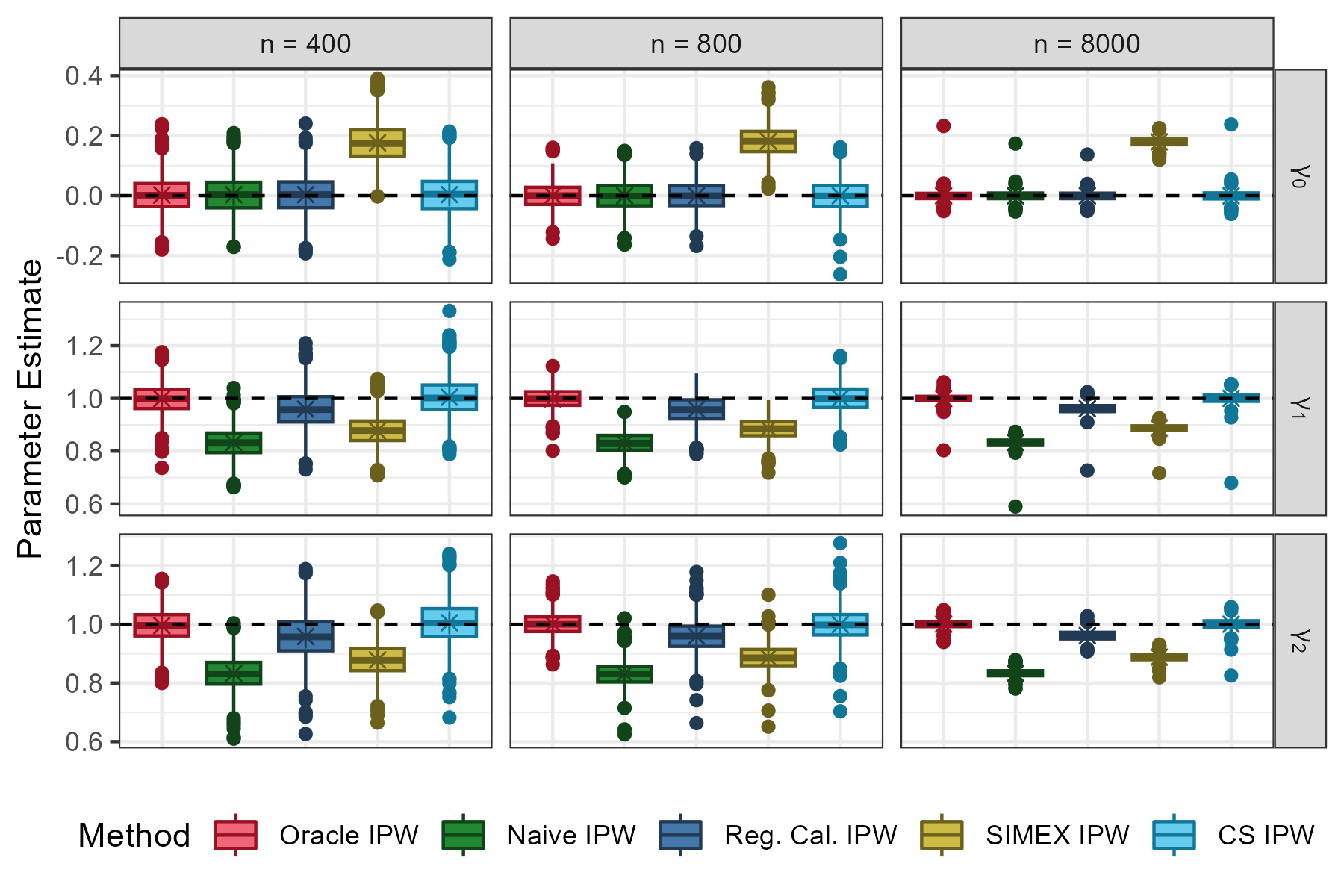}
\caption{Empirical distribution of MSM parameter estimates from the second simulation study, using the oracle, naive, regression calibration (reg. cal.), simulation-extrapolation (SIMEX), and corrected score (CS) IPW estimators.}
\label{fig:ipw}
\end{figure}

The MSM parameter of interest $\bgam = (\gamma_0, \gamma_{1}, \gamma_{2})$ was estimated for each simulated data set using five methods: (i) oracle IPW using the true values $A$, (ii) naive IPW treating the $A^*$ as perfectly measured, (iii) regression calibration IPW, (iv) SIMEX IPW, and (v) the proposed CS IPW estimator. Figure \ref{fig:ipw} shows the empirical distribution of MSM parameter estimates for the five methods and by sample size.

The oracle IPW estimator, which again serves as a benchmark, was unbiased. The naive and regression calibration IPW estimators of the intercept $\gamma_0$ had low bias, but the naive and regression calibration estimators of $\gamma_1$ and $\gamma_2$ had large bias that persisted with increasing sample size. The SIMEX IPW estimators of each parameter exhibited substantial bias. The CS IPW estimator on the other hand was essentially unbiased for each of the MSM parameters and across all sample sizes. The performance of standard error estimators and coverage of confidence intervals are shown in Web Appendix F.

The third simulation study compared the CS g-formula and IPW approaches to the CS DR estimator under various model specifications. In particular, 1000 datasets of $n \in \{400, 2000\}$ individuals were simulated with the following variables: confounders $L_{1} \sim \text{Bernoulli}(0.5)$ and $L_{2} \sim \sN(0, 0.16)$, exposure $A$ with $A|\bL \sim \sN(0.1 - 0.1L_1 + 0.3L_2, 0.04)$, and binary outcome $Y \sim \textrm{Bernoulli}(0.35 + 0.15A + 0.25L_1 + 0.2L_2 + 0.05AL_1 + 0.1AL_2)$ such that the assumptions of all three methods hold given correct model specifications. For this data generating process, the corresponding MSM was $\eta(a;\bgam) = \gamma_0 + \gamma_1 a$, where $\bgam = (\gamma_0, \gamma_1) = (0.475, 0.175$). The methods were compared with respect to their performance estimating $\gamma_1$. For the IPW estimator, this parameter was estimated directly. For the g-formula and DR estimators, this parameter was indirectly estimated as $\widehat\eta(1) - \widehat\eta(0)$; a corresponding variance estimator was constructed using the delta method. The exposure $A$ was subject to classical additive measurement error with known variance $\sigma_{me}^2 = 0.02$, corresponding to an exposure reliability of $0.69$. The three approaches were compared under scenarios where only the propensity model was correctly specified, only the outcome regression was correctly specified, both were correctly specified, or neither was correctly specified. The propensity model was mis-specified by not including the confounder $L_{1}$ and the outcome regression was mis-specified by leaving out $L_{1}$ and the interaction between $A$ and $L_{1}$.

\begin{table}[]
\fontsize{10.5}{12}\selectfont % Slightly smaller font (10.5pt size with 12pt line spacing)
    \centering
    %\footnotesize
    \caption{Results from the third simulation study. UC: uncorrected empirical sandwich variance estimator, BC: bias-corrected empirical sandwich variance estimator, n: sample size; Bias: 100 times the average bias across simulated data sets for each method; ESE: 100 times the standard deviation of parameter estimates; ASE: 100 times the average of estimated standard errors; Cov: Empirical percent coverage of 95$\%$ confidence intervals for each method. PS indicates the propensity score model is correctly specified; OR indicates the outcome regression is correctly specified.}
\begin{tabular}[t]{cccrrrrrr}
\toprule
\multicolumn{5}{c}{ } & \multicolumn{2}{c}{\textbf{UC}} & \multicolumn{2}{c}{\textbf{BC}} \\
\cmidrule(l{3pt}r{3pt}){6-7} \cmidrule(l{3pt}r{3pt}){8-9}
\textbf{n} & \textbf{Correct Specifications} & \textbf{Method} & \textbf{Bias} & \textbf{ESE} & \textbf{ASE} & \textbf{Cov} & \textbf{ASE} & \textbf{Cov}\\
 \midrule
 400 & PS and OR & CS DR & -0.1 & 18.7 & 16.8 & 91.5 & 17.0 & 92.1\\
  &  & CS G-Formula & 0.1 & 15.6 & 15.1 & 93.9 & 15.2 & 93.9\\
  &  & CS IPW & -0.7 & 19.3 & 17.6 & 92.1 & 18.0 & 92.2\\
 \addlinespace
  & PS Only & CS DR & -1.1 & 18.7 & 17.1 & 92.1 & 17.3 & 92.4\\
  &  & CS G-Formula & -14.9 & 15.0 & 14.9 & 82.8 & 14.9 & 83.0\\
  &  & CS IPW & -0.7 & 19.3 & 17.6 & 92.1 & 18.0 & 92.2\\
 \addlinespace
  & OR Only & CS DR & -0.4 & 16.9 & 15.9 & 93.1 & 16.1 & 93.1\\
  &  & CS G-Formula & 0.1 & 15.6 & 15.1 & 93.9 & 15.2 & 93.9\\
  &  & CS IPW & -15.3 & 16.5 & 16.0 & 83.3 & 16.2 & 83.9\\
 \addlinespace
  & Neither & CS DR & -15.3 & 16.4 & 15.7 & 81.7 & 15.9 & 82.2\\
  &  & CS G-Formula & -14.9 & 15.0 & 14.9 & 82.8 & 14.9 & 83.0\\
  &  & CS IPW & -15.3 & 16.5 & 16.0 & 83.3 & 16.2 & 83.9\\
 \addlinespace
 2000 & PS and OR & CS DR & 0.2 & 8.5 & 7.9 & 92.6 & 7.9 & 92.9\\
  &  & CS G-Formula & 0.0 & 6.8 & 6.6 & 94.2 & 6.6 & 94.3\\
  &  & CS IPW & 0.1 & 8.9 & 8.3 & 93.5 & 8.4 & 94.0\\
 \addlinespace
  & PS Only & CS DR & -0.1 & 8.6 & 8.0 & 93.3 & 8.1 & 93.4\\
  &  & CS G-Formula & -14.9 & 6.7 & 6.6 & 39.7 & 6.6 & 40.0\\
  &  & CS IPW & 0.1 & 8.9 & 8.3 & 93.5 & 8.4 & 94.0\\
 \addlinespace
  & OR Only & CS DR & 0.1 & 7.7 & 7.2 & 93.0 & 7.2 & 93.0\\
  &  & CS G-Formula & 0.0 & 6.8 & 6.6 & 94.2 & 6.6 & 94.3\\
  &  & CS IPW & -14.7 & 7.6 & 7.3 & 47.9 & 7.3 & 47.9\\
 \addlinespace
  & Neither & CS DR & -14.7 & 7.5 & 7.1 & 47.0 & 7.2 & 47.3\\
  &  & CS G-Formula & -14.9 & 6.7 & 6.6 & 39.7 & 6.6 & 40.0\\
  &  & CS IPW & -14.7 & 7.6 & 7.3 & 47.9 & 7.3 & 47.9\\
 \bottomrule
\end{tabular}
\label{tab:three}
\end{table}

The simulation results are presented in Table~\ref{tab:three} and support the theoretical results described in Section~\ref{s:dr}. Namely, when only the propensity score model was specified correctly, the IPW and CS DR estimators performed well, but the CS g-formula estimator was subject to substantial bias, and the corresponding CIs had lower than nominal coverage. Likewise when only the outcome model was specified correctly, the CS g-formula and DR estimators performed well, but the CS IPW estimator was biased, and the corresponding CIs had low coverage. That is, the CS DR estimator performed well when either one of the two models was mis-specified, exhibiting its namesake double-robustness property. When both models were correctly specified, the CS DR estimator performed similarly to the CS IPW estimator; in this scenario, the CS g-formula estimator had smaller variance than the other two methods. In general, the BC CIs tended to have slightly better coverage than the UC CIs in settings where the corresponding estimator was consistent.

Additional simulations of the proposed methods with estimated measurement error covariance, varied exposure reliability, a near positivity violation, and multiplicative measurement error covariance are presented in Web Appendix F.

\section{Application}
\label{s:application}

To illustrate the proposed methods, the CS DR estimator was applied to data from the HVTN 505 vaccine trial. This trial studied a candidate HIV vaccine with a primary endpoint of diagnosis of HIV-1 infection after the Month 7 study visit and through the Month 24 study visit. Immunologic markers were measured from blood samples at the Month 7 study visit. As discussed in the Introduction, the candidate HIV vaccine was not found to be effective, but follow-up research described several interesting immunologic marker correlates of risk. In particular, \citet{neidich2019} found that the immunologic markers (antibody-dependent cellular phagocytosis and antigen-specific recruitment of Fc$\gamma$ receptors of several HIV-1 specific Env proteins) were associated with reduced HIV-1 risk.

In this section, the primary analysis of \citet{neidich2019} is reassessed by (i) adjusting for measured potential confounders to target marginal effects and (ii) allowing for additive measurement error of the exposures. Analyses are done using the log transforms of markers measuring antibody-dependent cellular phagocytosis and recruitment of Fc$\gamma$R\RNum{2}a of the H131-Con S gp140 protein, which will be referred to as ADCP and R\RNum{2}. The primary analysis of \citet{neidich2019} focused on the association of each of these exposures individually with HIV-1 acquisition among vaccine recipients. For each exposure, \citet{neidich2019} fit a logistic regression model and reported odds ratios for the main effect of exposure adjusting for age, BMI, race, and behavior risk, as well as CD4 and CD8 polyfunctionality scores (CD4-P and CD8-P).

In this application, data are analyzed under a log link, i.e., a log-linear risk model, such that the proposed CS doubly-robust estimator can be used. A supplemental analysis using the CS g-formula estimator for an outcome model that is quadratic in the exposure is included in Web Appendix G. Notably, the immunologic markers were not measured in all participants, but rather were measured in all vaccine recipients with HIV acquisition and in a stratified random sample of vaccine recipients who completed 24 months of follow-up without HIV acquisition. To account for this two-phase sampling design, the weights in the doubly-robust estimator are multiplied by inverse probability of sampling weights, following the procedure described in \citet{wang2009}. Additionally, sampling weights were used to fit propensity score models in accordance with \citet{mccaffrey_estimating_2024}. This version of the proposed estimator is described in more detail and evaluated in a simulation study in Web Appendix H. ADCP and R\RNum{2} were modeled separately to match the univariate-style analysis performed in \citet{neidich2019}; accordingly, separate propensity models were fit for each exposure. Thus, the results cannot be interpreted as the joint effect of the two exposures.

For each propensity model specification, main effects for the covariates age, race, BMI, behavior risk, CD4-P, and CD8-P were included. For the outcome model specification, main effects for the exposure of interest, age, race, BMI, behavior risk, CD4-P, and CD8-P were included. Inverse probability of sampling weights were computed based on the case-cohort sampling design, with different weights estimated for cases and non-cases. The weight denominators were estimated with sample proportions (i.e., for non-cases, using the proportion of non-cases who had immunologic markers measured at Month 7). Based on the theoretical and empirical results in Sections~\ref{s:methods} and~\ref{s:simulation}, the doubly-robust CS estimator should be consistent if either specification is correct. Model diagnostics for the propensity and outcome models are presented in Web Appendix G. Finally, each exposure was assumed to follow a classical additive measurement error model where a sensitivity analysis was performed by varying the measurement error variances within 0, 1/8, 3/16, and 1/4 times the variance for each exposure variable when restricted to vaccine recipients with an immunologic marker. Since ADCP and R\RNum{2} are log-transformations of strictly positive random variables, this setup is equivalent to assuming that their corresponding non-log transformed variables follow multiplicative measurement error models. All covariates were assumed to be measured without error. Dose response curves were estimated for each exposure and each measurement error level across a range of 0.5 to 3 for ADCP and 7 to 10 for R\RNum{2}; exposure levels above 3 and 10 respectively were associated with no or almost no risk.

\begin{figure}[h!]
\centering
\includegraphics[width=6.5in]{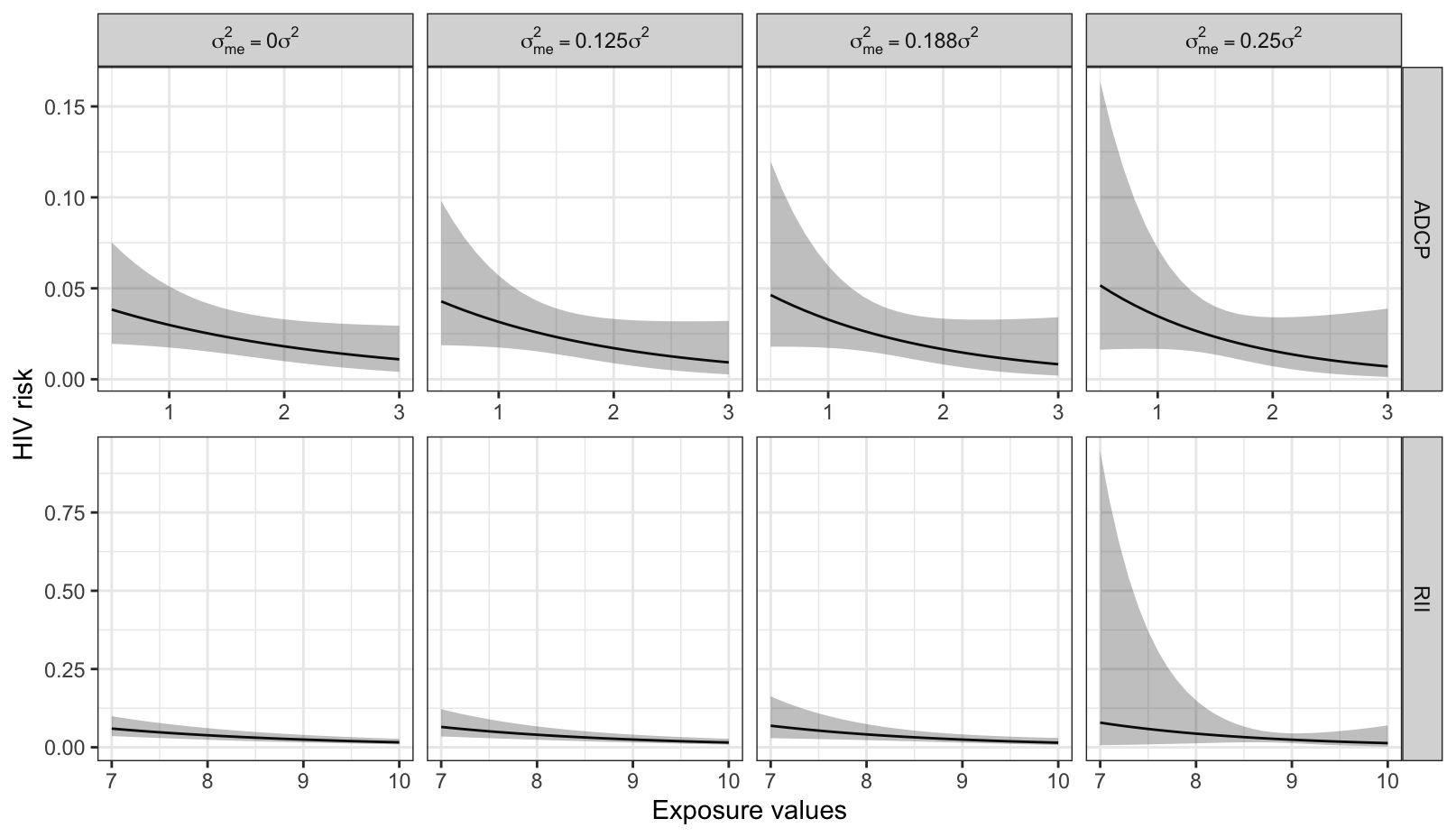}
\caption{HVTN 505 results. Each panel shows the dose-response curve for ADCP or R\RNum{2} estimated by the CS DR estimator, as well as their respective shaded 95\% pointwise confidence regions. From top to bottom, each panel reflects increasing user-specified variances of measurement error $\sigma^{2}_{me}$ corresponding to proportions of 0, 1/8, 3/16, and 1/4 of $\sigma^{2}$, the estimated total exposure variances among vaccinees with immune responses.}
\label{fig:three}
\end{figure}

The analysis results are plotted in Figure~\ref{fig:three}. For each exposure, lower values corresponded to higher HIV risk among the vaccine recipients, in line with prior results and biological theory. Moving across panels from left to right in Figure~\ref{fig:three}, the assumed measurement error variance increases and the confidence regions become wider, as expected. While higher levels of each exposure appear to cause lower risk, there is substantial uncertainty when allowing for higher magnitudes of measurement error and because of the low sample size (only 25 acquisition cases with measured Month 7 exposures). Thus, studies measuring these biomarkers in more participants are needed to draw stronger conclusions.

\section{Discussion}
\label{s:discussion}

In this paper, methods are proposed which adjust for both confounding and continuous exposure measurement error. The corrected score method, under an additive measurement error assumption, is applied to g-formula, IPW, and doubly-robust estimators. Empirical results confirm the results expected based on the large sample properties of the estimators. Accompanying this paper, an R package \texttt{mismex} has been developed.

While the proposed methods were shown to have favorable theoretical and empirical properties, they are not without limitation. In particular, the methods require that the measurement error covariance is known or can be estimated using data from previous studies or the current study. As demonstrated in Section~\ref{s:application}, if the covariance is unknown then a sensitivity analysis can be straightforward and informative if the covariance matrix is small or restricted such that it has few parameters. The proposed methods also rely on the additive measurement error assumption and correct specification of certain parametric models. The CS DR estimator provides some protection from mis-specification of outcome and propensity score models, requiring only one but not necessarily both to be correctly specified. When both models are mis-specified, DR estimators may perform worse than corresponding singly robust estimators~\citep{kang2007}, although this was not observed in the simulation study presented in Section~\ref{s:simulation}. Regardless of the estimator used, standard model-checking procedures should be employed in practice. A limitation of the data illustration in Section~\ref{s:application} was the assumption of no measurement error in covariates, which may not hold, e.g., for BMI. Additionally, if the two markers investigated in this application are immunologically linked, then the causal estimands targeted by the two individual analyses may be poorly defined or difficult to interpret since changing one marker would necessarily change the other.

There are several possible extensions of the methods described in this paper. The proposed methods only account for continuous exposures with classical additive measurement error and use cross-sectional data. Similar approaches could be developed that handle mis-classification of categorical exposures, use other measurement error models for continuous exposures, or allow time-varying exposures or survival outcomes. Developing approaches which adjust for covariate measurement error (as in \citealp{kuroki2014}; \citealp{lockwood2015}; \citealp{kyle2016}; \citealp*{hong2017}) along with exposure measurement error would be another useful extension. Yet another future direction of research is to combine g-estimation and CS methods to estimate parameters of a structural nested model where at least one exposure is measured with error. In addition, while this paper considers corrected score estimation methods, the conditional score estimation procedure \citep{carroll2006} is a related method which could be extended to a causal inference setting in similar ways. 

A final important direction of future research is to apply the CS methodology to efficient score functions. In semiparametric models, efficient score functions are estimating equations that produce estimators with maximum efficiency among all regular, asymptotically linear estimators \citep{kennedy_semiparametric_2015, tsiatis2006semiparametric}. Efficient score functions are widely used in causal inference due to their ability to handle complex or high-dimensional models for $Y|\bA,\bL$ or $\bA|\bL$. Since the CS approach in this paper can be applied to any estimating equation that is conditionally unbiased given $\bA$, if it can be shown that the efficient score function for a given statistical model satisfies this property, the CS method could be applied to allow for more flexible and efficient causal inference in the presence of a mismeasured exposure.

%  The \backmatter command formats the subsequent headings so that they
%  are in the journal style.  Please keep this command in your document
%  in this position, right after the final section of the main part of
%  the paper and right before the Acknowledgements, Supplementary Materials,
%  and References sections.

%\backmatter

%  This section is optional.  Here is where you will want to cite
%  grants, people who helped with the paper, etc.  But keep it short!

\section*{Acknowledgements}

The authors thank the Editor, Associate Editor, and two reviewers, as well as Kayla Kilpatrick, Shaina Mitchell, Sam Rosin, Bonnie Shook-Sa, and Jaffer Zaidi for helpful comments and suggestions. The authors also thank the investigators, staff, and participants of the HVTN 505 trial.

Research reported in this publication was supported by U.S. Public Health Service Grant AI068635, the National Institute of Allergy and Infectious Diseases of the National Institutes of Health (NIH) under Award Number R37 AI054165, the National Institute of Environmental Health Sciences of the NIH under Award Number T32 ES007018, and NIH grant R37AI029168. The content is solely the responsibility of the authors and does not necessarily represent the official views of the NIH.

\vspace*{-8pt}

\section*{Data Availability Statement}

The data that support the findings of this study are available on Atlas at \\ \href{https://atlas.scharp.org/cpas/project/HVTN\%20Public\%20Data/HVTN\%20505/begin.view}{https://atlas.scharp.org/cpas/project/HVTN\%20Public\%20Data/HVTN\%20505/begin.view}.\vspace*{-8pt}

%  Here, we create the bibliographic entries manually, following the
%  journal style.  If you use this method or use natbib, PLEASE PAY
%  CAREFUL ATTENTION TO THE BIBLIOGRAPHIC STYLE IN A RECENT ISSUE OF
%  THE JOURNAL AND FOLLOW IT!  Failure to follow stylistic conventions
%  just lengthens the time spend copyediting your paper and hence its
%  position in the publication queue should it be accepted.

%  We greatly prefer that you incorporate the references for your
%  article into the body of the article as we have done here
%  (you can use natbib or not as you choose) than use BiBTeX,
%  so that your article is self-contained in one file.
%  If you do use BiBTeX, please use the .bst file that comes with
%  the distribution.

\bibliographystyle{biom} 
\bibliography{references}

\section*{Supporting Information}

Web Appendices, Tables, and Figures referenced in Sections~\ref{s:methods},~\ref{s:simulation}, and~\ref{s:application} are  available at the end of this document. All R code used in the simulations and application is available at the Biometrics website or in the R package \texttt{mismex} at \href{https://github.com/brian-d-richardson/mismex}{https://github.com/brian-d-richardson/mismex}.\vspace*{-8pt}

\newpage
\thispagestyle{empty} % Remove header and footer from this page

\appendixtitle

\newpage

% Commands specific to appendix formatting
% Reset section counter for appendix
\setcounter{section}{0}

% Reset table counter and redefine numbering
\setcounter{table}{0}
\renewcommand{\thetable}{\arabic{table}}
\makeatletter
\renewcommand{\fnum@table}{Web Table \thetable}
\makeatother

% Reset figure counter and redefine numbering
\setcounter{figure}{0}
\renewcommand{\thefigure}{\arabic{figure}}
\makeatletter
\renewcommand{\fnum@figure}{\textbf{Web Figure \thefigure}}
\makeatother

% Reset section numbering format for appendix
\renewcommand{\thesection}{\Alph{section}}

\section{Fitting propensity score models}

In this section, details for fitting propensity score models are discussed. Methods for fitting MSMs with multiple treatments often use weights of the form $SW = \prod_{j=1}^{m} f_{A_j}(A_{j}) / f_{A_j|\bL}(A_{j} | \bL)$, e.g., as in \citet{hernan2001}; to factorize the denominator in this way, the $m$ exposures $A_{1}, \ldots, A_{m}$ are assumed to be independent conditional on $\bL$. This assumption, which is made in the simulation studies, allows one to avoid multivariate conditional density estimation, but it may be dubious in various applications, such as when a treatment has a direct effect on another treatment or when treatments have an unmeasured common cause. This assumption does have testable implications; see \citet{zhang2012} for a related testing procedure. Alternatively, the denominator of $SW$ could be estimated using a mixed effects model \citep{tchetgen2012}.

Models used to estimate weights can be fit using various previously described approaches. For simplicity, in the remainder of this paper, weight components corresponding to continuous exposures will be constructed using a ratio of normal densities estimated from linear models~\citep{hirano2004}. To illustrate this, first consider the setting where the true exposures are observed. For continuous exposure $A_{j}$, a model of the form $A_{j} = \bzeta (1, \bL)^{T} + \epsilon_{ps}$ might be used, where $\epsilon_{ps} \sim \mathcal{N}(0, \sigma^{2}_{ps})$. Then based on the fitted model, the estimated conditional density $f_{\bA_j|\bL}(A_{j} | \bL; \widehat{\bzeta})$ is used to estimate $f_{\bA_j|\bL}(A_{j} | \bL)$. An intercept-only model is used similarly to estimate the weight numerators. Other methods and more flexible choices for weight models~\citep{naimi2014} can also be used for continuous exposures in practice.

When a parametric propensity model is used for the denominator of $SW$, integrating the assumed $\bA | \bL$ model over the covariate distribution implies certain restrictions on the marginal distribution of $\bA$. However, the consistency of the IPW estimator does not rely on a correctly specified model for the numerator of $SW$ (see, e.g., Technical Point 12.2 of \cite{hernan2020}), such that the numerator and denominator models do not necessarily have to be compatible. Finally, weight components corresponding to discrete exposures (which are assumed to be always correctly measured) can be estimated using common approaches, such as logistic and multinomial regression.

A correctly specified model is required in order for the estimated $f_{\bA|\bL}$ to be consistent. Even in the case where the exposure is perfectly measured, this assumption may be dubious; this limitation motivates the DR estimator in Section 3.4 of the main text. When the exposure is mismeasured, consistency of $f_{\bA|\bL}$ relies on the additional assumptions that $\beps_{me} \perp \!\!\! \perp Y, \bA, \bL$ and that $\bSigma_{me}$ is known. These assumptions, which are required for all of the estimators proposed in the main text, should be assessed carefully.

In the case where only a mismeasured exposure is observed, the densities $f_{\bA}(\bA; \bxi)$ and $f_{\bA|\bL}(\bA | \bL; \bzeta)$ for the true exposure $\bA$ using the measured $\bAs$ can be estimated using existing methods (e.g., regression calibration or simulation-extrapolation). If $\bA | \bL$ follow a multivariate normal distribution with conditionally independent outcomes and a linear mean model, the estimating functions corresponding to ordinary least squares (OLS) with $\bAs$ in place of $\bA$ will give consistent estimators of the mean model coefficients, and the usual OLS covariance estimator is consistent for $\Cov(\bAs|\bL) = \Cov(\bA|\bL) + \bSigma_{me}$, from which $\Cov(\bA|\bL)$ can be recovered since $\bSigma_{me}$ is assumed known.

\section{Closed form corrected score functions}
\label{sec:closed_forms}

In Section 3.5 of the main text, three examples of corrected score functions with known closed forms are mentioned. Here, a closed form for the CS IPW function is provided in the case where the PS models for $\bA$ and $\bA | \bL$ are normal with identity links and where the MSM has a linear link. The same strategy can be used to find a closed form for the DR CS function with the same type of PS models, and where the outcome model has an identity link.

For the sake of simplicity, consider a univariate exposure $A$ subject to measurement error $\epsilon \sim \mathcal{N}(0,\sigma_{me}^2)$, where $A|\bL \sim \mathcal{N}(\mu_{\bL}, \delta^2)$, $A \sim \mathcal{N}(\mu,\tau^2)$, and the PS model parameters are known. This result can generalize to multidimensional $\bA$ with parameters estimated using the root of $\bPsi_{PS}(\bL, \bA ; \bxi, \bzeta)$. In this setting and in the absence of measurement error, the IPW estimating function is
\begin{align*}
    \bPsi_{0-IPW}(Y, \bL, A ; \btheta_{IPW}) =
       SW(\bL, A)
       \left\{ Y - (\gamma_0 - \gamma_a A) \right\}
       (1, A),
\end{align*}
where, using the normality of $A$ and $A|\bL$, 
\begin{align*}
    SW(\bL, A) &=
    \frac{\delta}{\tau}\exp \left\{
    \underbrace{\frac{1}{2}\left(\delta^{-2} - \tau^{-2}\right)}_{b_1} A^2 +
    \underbrace{\left(\frac{\mu}{\tau^2} - \frac{\mu_{\bL}}{\delta^2}\right)}_{b_2} A +
    \underbrace{\frac{1}{2}\left(\frac{\mu^2}{\tau^2} - \frac{\mu_{\bL}^2}{\delta^2}\right)}_{b_3}
    \right\} \\
    &= \frac{\delta}{\tau}\exp \left(
    b_1 A^2 + b_2 A + b_3 \right).
\end{align*}
Then the CS IPW estimating function is $\bPsi_{CS-IPW}(Y, \bL, \As ; \sigma^2_{me}, \btheta_{IPW})$
\begin{align}
\label{eq:csipw1}
    &=\E \left( \real \left[ SW(\bL, \At)
       \left\{ Y - (\gamma_0 + \gamma_1 \At) \right\}
       (1, \At)
       \right] | Y, \bL, \As \right) \nonumber \\
    &=\E \left\{ \real \left( SW(\bL, \As + i\epst)
       \left[ Y - \{\gamma_0 + \gamma_1 (\As + i\epst)\} \right]
       (1, \As + i\epst)
       \right) | Y, \bL, \As \right\}.
\end{align}
Note that the expectation in \eqref{eq:csipw1} is conditional on $Y, \bL, \As$, and the only random component is $\epst$. With this in mind, expressions in \eqref{eq:csipw1} can be written as
\begin{align*}
    SW(\bL, \As + i\epst) &= \frac{\delta}{\tau}
    \exp\left\{
    \underbrace{b_1 A^{*2} + b_2 \As + b_3}_{c_1} -
    \underbrace{b_1}_{c_2} \epst^2 +
    \underbrace{(2b_1 + b_2)}_{c_3} i \epst
    \right\} \\
    &= \frac{\delta}{\tau} \exp\left(
    c_1 - c_2 \epst^2 + c_3 i \epst
    \right),
\end{align*}
and
\begin{align*}
    \left[ Y - \{\gamma_0 - (\As + i \epst)\} \right]
    \begin{bmatrix}
    1 \\ \As + i\epst
    \end{bmatrix}^{T} &= 
    \begin{bmatrix}
    \overbrace{Y - \gamma_0 - \gamma_1 \As}^{d_1} -
    \overbrace{\gamma_2}^{d_2} i \epst  \\
    \underbrace{(Y - \gamma_0) - \gamma_1 A^{*2}}_{d_3} +
    \underbrace{\gamma_1}_{d_4} \epst^2 +
    \underbrace{(Y - \gamma_0 - 2\gamma_1 \As)}_{d_5} i \epst
    \end{bmatrix}^{T} \\
    &= \begin{bmatrix}
    d_1 - d_2 i \epst  \\
    d_3 + d_4 \epst^2 + d_5 i \epst
    \end{bmatrix}^{T},
\end{align*}
where $c_1, c_2, c_3, d_1, d_2, d_3, d_4, d_5$ are real constants with respect to the conditional expectation. Then \eqref{eq:csipw1} can be rewritten as
\begin{align}
\label{eq:csipw2}
    \frac{\delta}{\tau} \E \left[ \real \left\{
    \exp \left( c_1 - c_2 \epst^2 + c_3 i \epst \right)
    \begin{bmatrix}
    d_1 - d_2 i \epst  \\
    d_3 + d_4 \epst^2 + d_5 i \epst
    \end{bmatrix}^{T}
    \right\} \right].
\end{align}
The expectation in \eqref{eq:csipw2} can be evaluated using Euler's formula:
\begin{align*}
    \exp(i \alpha) &= \cos(\alpha) + i \sin(\alpha) \hspace{1em} \forall \alpha \in \mathbb{R},
\end{align*}
and the characteristic function of $\epst$:
\begin{align*}
    \E \{\exp(i \epst) \} &= \exp(-\sigma_{me}^2 / 2),
\end{align*}
to show that $\bPsi_{CS-IPW}(Y, \bL, \As ; \sigma^2_{me}, \btheta_{IPW})$ 
\begin{multline*}
    = \frac{\delta}{\tau}
    (1 - 2 \sigma_{me}^2 c_2) ^ {-1/2}
    \exp\left\{ c_1 - \frac{c_3 ^ 2}{2 (\sigma_{me}^{-2} - 2 c_2)} \right\} \\
    \times
    \begin{bmatrix}
        d_1 + d_2 c_3(\sigma_{me}^{-2} - 2 c_2)^{-1} \\
        d_3 - d_5 c_3(\sigma_{me}^{-2} - 2 c_2)^{-1} +
        d_4(\sigma_{me}^{-2} - 2c_2)^{-1}\left(1 - \frac{c_3^2}{\sigma_{me}^{-2} - 2c_2}\right)
    \end{bmatrix}^{T}.
\end{multline*}

\section{Large sample properties}
\label{sec:asymp-props}

In this appendix, the large sample properties of the proposed estimators discussed in Section 3 of the main paper are proven. In particular, the g-formula, IPW, and doubly-robust estimating functions $\bPsi_{0-GF}, \bPsi_{0-IPW}, \bPsi_{0-DR}$ which ignore measurement error are shown to be unbiased (i.e., have expected value equal to $\bo$) and the components that involve $\bA$ are shown to be conditionally unbiased given $\bA$. Given these results, it follows that the three proposed corrected score estimators are consistent and asymptotically normal \citep{carroll2006}.

\subsection{G-formula CS estimator}
\label{sec:gfmla-props}

Let $\bPsi^{(k)}$ denote the $k^{\rm th}$ component of an estimating function $\bPsi$. Consider the conditional expectation of the first component of the g-formula estimating function,
\begin{align*}
    \E\left\{ \bPsi_{0-GF}^{(1)}(Y,\bL,\bA;\btheta_{GF}) | \bA \right\}
    &= \E \left[ \left\{ Y - \mu(\bL, \bA; \bbeta) \right\} \pd{\bbeta}^{T} \mu(\bL, \bA; \bbeta) | \bA \right] \\
    &= \E \left( \E \left[  \left\{ Y - \mu(\bL, \bA; \bbeta) \right\}
         \pd{\bbeta}^{T} \mu(\bL, \bA; \bbeta) | \bL, \bA \right] | \bA \right) \\
    &= \E \left[ \left\{ \underbrace{\E (Y | \bL, \bA) - \mu(\bL, \bA; \bbeta)}_{=0} \right\}
         \pd{\bbeta}^{T} \mu(\bL, \bA; \bbeta) | \bA \right] \\   
    &= \bo,
\end{align*}
where the last inequality follows by the assumption of a correctly specified outcome model. The second component of $\bPsi_{0-GF}$, which does not involve $\bA$, has marginal expectation
\begin{align*}
    \E\left\{ \bPsi_{0-GF}^{(2)}(Y,\bL,\bA;\btheta_{GF}) \right\}
    &= \E \left\{ \eta(\ba) - \mu(\bL, \ba; \bbeta) \right\} \\
    &= \eta(\ba) - \E \left\{ \mu(\bL, \ba; \bbeta) \right\} \\
    &= \eta(\ba) - \E \left\{ \E (Y | \bL, \bA = \ba) \right\}
        && \text{(correct specification of } \mu \text{)} \\
    &= \eta(\ba) - \E \left[ \E \{Y(\ba) | \bL, \bA = \ba\} \right]
        && \text{(causal consistency)} \\
    &= \eta(\ba) - \E \left[ \E \{Y(\ba) | \bL \} \right]
        && \text{(conditional exchangeability)} \\
    &= \eta(\ba) - \E \{ Y(\ba) \} \\
    &= \bo.
\end{align*}

\subsection{CS IPW estimator}
\label{sec:ipw-props}

For the IPW estimator, it is assumed that the parameters $\bxi, \bzeta$ in the propensity models for $\bA$ and $\bA | \bL$ are consistently estimated with the root of $\sum_{i=1}^n\bPsit_{PS}(\bL_i, \bAs_i; \bSigma_{me}, \bxi, \bzeta)$. Then it remains to show that the second component of $\bPsi_{0-IPW}$ is conditionally unbiased.
\begin{align*}
    &\E\left\{ \bPsi_{0-IPW}^{(2)}(Y,\bL,\bA;\btheta_{IPW}) | \bA = \ba \right\} \\
    &= \E \left[ SW(\bL, \bA)
       \left\{ Y - \eta(\bA; \bgam) \right\} \pd{\bgam}^{T} \eta(\bA; \bgam) | \bA = \ba \right] \\
    &= \E \left[ \frac{ f_{\bA}(\bA; \bxi) }{ f_{\bA | \bL}(\bA | \bL)}
       \left\{ Y - \eta(\bA; \bgam) \right\} \pd{\bgam}^{T} \eta(\bA; \bgam) | \bA = \ba \right] \\
    &= f_{\bA}(\ba; \bxi) \pd{\bgam}^{T} \eta(\ba; \bgam)
        \E \left\{ \frac{ Y(\ba) - \eta(\ba; \bgam) }{ f_{\bA | \bL} (\ba | \bL)}
        | \bA = \ba \right\} \\
    &= f_{\bA}(\ba; \bxi) \pd{\bgam}^{T} \eta(\ba; \bgam)
        \int \frac{ \E\{Y(\ba) | \bA = \ba, \bL = \bl \} -
        \eta(\ba; \bgam) }{ f_{\bA | \bL} (\ba | \bl)}
        f_{\bL | \bA} (\bl | \ba) d\bl \\
    &= f_{\bA}(\ba; \bxi) \pd{\bgam}^{T} \eta(\ba; \bgam) f_{\bA}(\ba)^{-1}
        \int \left[ \E\{Y(\ba) | \bL = \bl \} -
        \eta(\ba; \bgam) \right] f_{\bL} (\bl) d\bl \\  
    &= f_{\bA}(\ba; \bxi) \pd{\bgam}^{T} \eta(\ba; \bgam) f_{\bA}(\ba)^{-1}
        \left[ \E\{Y(\ba)\} -
        \eta(\ba; \bgam) \right] \\  
    &= \bo.
\end{align*}

Note that this conditional unbiasedness result does not rely on $f_{\bA}(\bA; \bxi)$ being the correct marginal density for $\bA$, but it does rely on $f_{\bA|\bL}(\bA|\bL; \bzeta)$ being the correct conditional density for $\bA | \bL$.

\subsection{Doubly robust CS estimator}
\label{sec:dr-props}

To show that the doubly robust estimator is in fact doubly robust, we show that $\bPsi_{CS-DR}$ is unbiased when either (i) the outcome model $\mu(\bL, \bA; \bbeta)$ or (ii) the propensity model $f_{\bA | \bL}(\bA, \bL; \bxi)$ is correctly specified. In the former case, we show that the second and third components of $\bPsi_{CS-DR}$ are unbiased at the true values of $\bbeta, \eta(\ba)$ and at arbitrary values of $\bxi, \bzeta$. In the latter case, we show that the first and third components of $\bPsi_{CS-DR}$ are unbiased at the true values of $\bxi, \bzeta, \eta(\ba)$ and at the root $\bbeta^*$ of the expected value of the second component. In both cases, the unbiasedness of the third component, which corresponds to the parameter of interest $\eta(\ba)$, implies consistent estimation of $\eta(\ba)$.

\subsubsection{Correctly specified outcome model}
\label{sec:dr-outcome}

Suppose the outcome model is correctly specified such that $\mu(\bL, \bA; \bbeta) = \E(Y | \bL, \bA)$ at the true value of $\bbeta$, and the propensity model is possibly misspecified such that the denominator of $SW(\bL, \bA)$ may not equal the true density $f_{\bA|\bL}(\bA|\bL)$. Then the second component of $\bPsi_{0-DR}$ has conditional expectation
\begin{align*}
    &\E \left[ SW(\bL, \bA)
          \{ Y - \mu(\bL, \bA; \bbeta) \}
          \pd{\bbeta}^{T} \mu(\bL, \bA; \bbeta) | \bA \right] \\
    &= \E \left( \E \left[ SW(\bL, \bA)
          \{ Y - \mu(\bL, \bA; \bbeta) \}
          \pd{\bbeta}^{T} \mu(\bL, \bA; \bbeta) | \bL, \bA \right] | \bA \right) \\
    &= \E \left( SW(\bL, \bA)
          \underbrace{ \{ \E(Y | \bL, \bA) - \mu(\bL, \bA; \bbeta) \} }_{=0}
          \pd{\bbeta}^{T} \mu(\bL, \bA; \bbeta) | \bA \right) \\
    &= \bo.
\end{align*}
The third component of $\bPsi_{0-DR}$ equals the second component of $\bPsi_{0-GF}$, which was shown in Web Appendix \ref{sec:gfmla-props} to be marginally unbiased when the outcome model is correctly specified.

\subsubsection{Correctly specified propensity model}
\label{sec:dr-propensity}

Now suppose the propensity model is correct but the outcome model may be incorrect. That is, $SW(\bL, \bA; \bxi, \bzeta) = h(\bA) / f_{\bA|\bL}(\bA | \bL)$ for some function $h$, but the posited $\mu(\bL, \bA; \bbeta)$ may not equal the true $\E(Y | \bL, \bA)$ for any value of $\bbeta$. Let $\bbeta^* = \{\beta^*_0, \bbeta^*_l, \bbeta^*_a, \vc(\bbeta^*_{al})\}$ be the root of $\E \{ \bPsi_{CS-DR}^{(2)}(Y, \bL, \bAs ; \bbeta) \}$. Then we claim that $\bPsi_{0-DR}$ is unbiased at $\btheta^{*}_{DR} = \{\bbeta^{*}, \eta(\ba; \bgam), \bxi, \bzeta\}$, where $\eta(\ba; \bgam)$ is the true MSM. The proof has two steps: (i) to show that $\bbeta^*$ is the root of $\E \left\{ \bPsi_{0-DR}^{(2)}(Y, \bL, \bA ; \bbeta) | Y, \bL \right\}$, and (ii) to show that this implies $\E\{ Y(\ba) \} = \E\{ \mu(\bL, \ba; \bbeta^*) \}$.

For the first step, we will use use equation (7.29) in Chapter 7 of \cite{carroll2006} which says that, for a suitably smooth and integrable function $g(\bA)$,
\begin{align}
\label{eq:caroll729}
    g(\bA) = \E \left( \E \left[ \real \left\{ g(\bAt) \right\} | Y, \bAs, \bL \right] | \bA \right).
\end{align}
Taking the expectation of both sides of \eqref{eq:caroll729} implies
\begin{align}
\label{eq:caroll729-2}
    \E \{ g(\bA) \} 
    &= \E \left\{ \E \left( \E \left[ \real \left\{ g(\bAt) \right\} | Y, \bAs, \bL \right] | \bA \right) \right\} \nonumber \\
    &= \E \left( \E \left[ \real \left\{ g(\bAt) \right\} | Y, \bAs, \bL \right] \right) \nonumber \\
    &= \E \left[ \real \left\{ g(\bAt) \right\} \right].
\end{align}
By the definition of $\bbeta^*$, and letting $g(\bA) = \bPsi_{0-DR}^{(2)}(Y, \bL, \bA ; \bbeta^{*})$,
\begin{align}
\label{eq:dr-step1}
    \bo &= \E \left\{ \bPsi_{CS-DR}^{(2)}(Y, \bL, \bA^* ; \bbeta^{*}) \right\} 
        && (\text{definition of } \beta^{*}) \nonumber \\
    &= \E \left[  \E \left\{ \bPsi_{CS-DR}^{(2)}(Y, \bL, \bA^*, \bbeta^{*})
        | Y, \bL \right\} \right]
        && (\text{iter. cond. exp.}) \nonumber \\
    &= \E \left[ \E \left( \E \left [ \real \left\{ 
        \bPsi_{0-DR}^{(2)}(Y, \bL, \bAt ; \bbeta^{*})
        \right \}  | Y, \bL, \bAs \right] 
        | Y, \bL \right) \right]
        && (\text{definition of } \bPsi_{CS-DR}^{(2)}) \nonumber \\
    &= \E \left( \underbrace{ \E \left [ \real \left\{ 
        \bPsi_{0-DR}^{(2)}(Y, \bL, \bAt ; \bbeta^{*})
        \right \}  | Y, \bL \right] }_{ 
        \E \left[ \real \left\{ g(\bAt) \right\} | Y, \bL \right] }
        \right) 
        && (\text{undo iter. cond. exp.}) \nonumber \\
    &= \E \left [ \underbrace{ \E \left\{ 
        \bPsi_{0-DR}^{(2)}(Y, \bL, \bA ; \bbeta^{*}) | Y, \bL
        \right\} }_{ \E\{ g(\bA) | Y, \bL \} }
        \right]
        && (\text{equation } \eqref{eq:caroll729-2}) \nonumber \\
    &= \E \left\{ \bPsi_{0-DR}^{(2)}(Y, \bL, \bA ; \bbeta^{*}) \right\}.
        && (\text{undo iter. cond. exp.})
\end{align}
The second step follows the proof in \cite{zhang2016} of double robustness of this estimator in the absence of measurement error. Specifically, \eqref{eq:dr-step1} implies
\begin{align*}
    \bo &= \E \left\{ \bPsi_{0-DR}^{(2)}(Y, \bL, \bA ; \bbeta^*) \right\} \\
        &= \E\left[ SW(\bL, \bA ; \bxi, \bzeta)
        \left\{ Y - \mu(\bL, \bA; \bbeta^*) \right\}
        \pd{\bbeta}^{T} \mu(\bL, \bA; \bbeta^*) \right],%  \\
\end{align*}
which in turn implies 
\begin{align}
\label{eq:dr1}
    \bo &=
    \E\left[ SW(\bL, \bA ; \bxi, \bzeta)
        \{ Y - \mu(\bL, \bA; \bbeta^*) \}
        (1, \bA) \right] \nonumber \\
    &= \E \left( \E\left[ SW(\bL, \bA ; \bxi, \bzeta)
        \{ Y - \mu(\bL, \bA; \bbeta^*) \}
        (1, \bA) | \bA \right] \right) \nonumber \\
    &= \E \left( h(\bA) \left[ 
        \E \left\{ \frac{ Y }{ f_{\bA | \bL}(\bA | \bL) } | \bA \right\} - 
        \E \left\{ \frac{ \mu(\bL, \bA; \bbeta^*) }{ f_{\bA | \bL}(\bA | \bL) } | \bA \right\}
        \right] (1, \bA) \right).
\end{align}
By arguments similar to those in Appendix \ref{sec:ipw-props}, it can be shown that
\begin{align*}
    h(\bA) E \left\{ \frac{ Y }{ f_{\bA | \bL}(\bA | \bL) } | \bA \right\} = 
    \eta(\bA; \bgam)
\end{align*}
and
\begin{align*}
    h(\bA) \E \left\{ \frac{ \mu(\bL, \bA; \bbeta^*) }
        { f_{\bA | \bL}(\bA | \bL) } | \bA \right\}
    &= \int \mu(\bl, \bA; \bbeta^*) f_{\bL}(\bl) d\bl \\ 
    &= \eta(\bA; \bgam^*),
\end{align*}
where $\bgam^*$ is obtained by replacing $\bbeta$ for $\bbeta^*$ in the expression for $\bgam$. Then substituting these expressions into \eqref{eq:dr1} gives
\begin{align*}
    \bo &= \E \left[ \left\{
        \eta(\bA; \bgam)  - 
        \eta(\bA; \bgam^*)
        \right\} (1, \bA) \right],
\end{align*}
which implies that $\bgam^* = \bgam$.
Thus,  $\E\{ Y(\ba) \} = \E\{ \mu(\bL, \ba; \bbeta^*) \}$ and $\bPsi_{CS-DR}$ is unbiased at $\btheta^*_{DR} = \{\bbeta^*, \eta(\ba; \bgam), \bxi, \bzeta\}$.

\section{Estimating measurement error covariance}
\label{s:estvar}

In this section, details are provided for how the measurement error covariance can be estimated using supplemental data. In some cases, supplemental data from the current study may be available in the form of replicates of the potentially mismeasured variables. These replicates can be used to estimate $\bSigma_{me}$ as described in \citet{carroll2006}. In particular, suppose for individual $i$ there are $k_{i}$ replicates of the mismeasured exposures, $\bAs_{i1}, ..., \bAs_{ik_{i}}$ with mean $\bAs_{i.}$. Then an estimator for the measurement error covariance is given by:
\begin{equation}
\label{eq:estvar}
    \bSigmahat_{me} = \frac{\sum_{i=1}^{n} \sum_{j=1}^{k_{i}} (\bAs_{ij} - \bAs_{i.})^{T}(\bAs_{ij} - \bAs_{i.})}{\sum_{i=1}^{n}(k_{i} - 1)}
\end{equation}
Likewise if validation data are available, the measurement error covariance can be estimated e.g., using maximum likelihood. There are many types of studies for which the covariance matrix may be assumed to follow a certain structure (e.g., diagonal, meaning the measurement errors are uncorrelated). For example, biological assays run on different samples and analyzed by separate machines/researchers may have uncorrelated or only weakly correlated measurement errors \citep{farrance_uncertainty_2012}. In these cases $\bSigmahat_{me}$ can be modified to fit the assumed structure. For other types of data such as survey responses, analysts should be more cautious, noting that response bias, recall bias, and other forms of measurement error in survey instruments may be correlated within individuals~\citep{biemer2013}. Regardless of how the measurement error covariance is estimated, the methods described in this paper can be used with $\bSigma_{me}$ replaced with $\bSigmahat_{me}$. Section 7.5.2 of \citet{carroll2006} provides one way to estimate the covariance of the the estimator $\bthetahat$, which accounts for uncertainty in $\bSigmahat_{me}$. Alternatively, provided $\bSigmahat_{me}$ is an M-estimator, the estimating equations for $\bSigmahat_{me}$ and $\bthetahat$ can be stacked and solved simultaneously, and the corresponding sandwich variance estimator will reflect uncertainty in $\bSigmahat_{me}$ \citep{cole_illustration_2023, shooksa_fusing_2024}.

\section{Regression calibration and simulation extrapolation}
\label{sec:rc-simex}

In the the first two simulation studies in Section 4 of the main text, the proposed CS estimators are compared to regression calibration and simulation extrapolation (SIMEX) estimators. In this section, details of these methods are provided.

\subsection{Regression calibration}
\label{sec:rc}

Regression calibration entails substituting an estimate $\Ehat(\bA|\bAs,\bL)$ of $\E(\bA|\bAs,\bL)$ for $\bA$ in an estimating equation \citep{carroll2006}. The exact form of this expectiation can be difficult to calculate, but a common approximation is given in Equation (4.4) from \citet{carroll2006}:
\begin{align}
    \E(\bA|\bAs,\bL) \approx \E(\bA) +
    \begin{bmatrix}
        \Cov(\bA) \\
        \Cov(\bL, \bA)
    \end{bmatrix}^T
    \begin{bmatrix}
        \Cov(\bA) + \bSigma_{me} & \Cov(\bA,\bL) \\
        \Cov(\bL, \bA) & \Cov(\bL)
    \end{bmatrix}^{-1}
    \begin{bmatrix}
        \bAs - \E(\bA) \\
        \bL - \E(\bL)
    \end{bmatrix}.
\end{align}
Then $\E(\bA|\bAs,\bL)$ can be estimated with
\begin{align}
    \Ehat(\bA|\bAs,\bL) =
    \Ehat(\bA) +
    \begin{bmatrix}
        \Covhat(\bA) \\
        \Covhat(\bL, \bA)
    \end{bmatrix}^T
    \begin{bmatrix}
        \Covhat(\bA) + \bSigma_{me} & \Covhat(\bA,\bL) \\
        \Covhat(\bL, \bA) & \Covhat(\bL)
    \end{bmatrix}^{-1}
    \begin{bmatrix}
        \bAs - \Ehat(\bA) \\
        \bL - \Ehat(\bL)
    \end{bmatrix},
\end{align}
where
\begin{align*}
    \Ehat(\bA) &= n^{-1}\sum_{i=1}^n\bAs_i, \\
    \Ehat(\bL) &= n^{-1}\sum_{i=1}^n\bL_i, \\
    \Covhat(\bA) &= (n-1)^{-1}\sum_{i=1}^n\{\bAs_i - \Ehat(\bA)\}^T\{\bAs_i - \Ehat(\bA)\} - \bSigma_{me}, \\
    \Covhat(\bA, \bL) &= (n-1)^{-1}\sum_{i=1}^n\{\bAs_i - \Ehat(\bA)\}^T\{\bL_i - \Ehat(\bL)\}, \\
    \Covhat(\bL) &= (n-1)^{-1}\sum_{i=1}^n\{\bL_i - \Ehat(\bL)\}^T\{\bL_i - \Ehat(\bL)\}.
\end{align*}

From here, a regression calibration version of the g-formula, IPW, or DR estimator can be constructed by substituting $\Ehat(\bA|\bAs,\bL)$ for $\bA$ in $\bPsi_{0-GF}$, $\bPsi_{0-IPW}$, or $\bPsi_{0-DR}$. This approach is used in the first two simulations of the main text.

\subsection{Simulation extrapolation}
\label{sec:simex}

Broadly, SIMEX entails adding varying amounts of simulated measurement error to the mismeasured exposure,  recalculating the estimator at each level of added error, then extrapolating backwards to obtain an estimate under no total measurement error \citep{carroll2006}. For a given estimating equation $\bPsi_0$ and estimand $\btheta$, let $\bthetahat_b(\lambda)$ be the estimate obtained by finding the root of $\sum_{i=1}^n \bPsi_0(Y_i, \bL_i, \bAs_i + \sqrt{\lambda}\beps_{b,i} ; \btheta)$, where $\beps_{b,i}$ are iid $\bsN(\bo,\bSigma_{me})$  simulated measurement errors, $i\in\{1,\dots,n\}$, and $b\in\{1,\dots,B\}$. Then let $\bthetahat(\lambda)=B^{-1}\sum_{b=1}^B\bthetahat_b(\lambda)$ be the average over $B$ replicates with the same $\lambda$. The SIMEX estimator of $\btheta$ is the extrapolation to $\bthetahat(-1)$. 

For the SIMEX g-formula estimator in the first simulation study of the main text, an estimator $\bbetahat_{SIMEX}$ of $\bbeta$ was obtained using the \texttt{simex} R package \citep{lederer_simex_2005} with $B=100$, $\lambda \in \{0.5, 1, 1.5, 2\}$, and using quadratic extrapolation. Then the dose response curve was estimated as $\etahat_{SIMEX}(\ba) = n^{-1} \sum_{i=1}^{n} \mu(\bL_i, \ba; \bbetahat_{SIMEX})$. For the SIMEX IPW estimator in the second simulation study of the main text, the \texttt{simex} package could not be used, and the simulation and extrapolation steps were coded manually using the same $B$ and set of $\lambda$, and using quadratic extrapolation.

\section{Additional simulations}
\label{sec:additional-sims}

In this section, additional simulations are presented to assess the estimated standard errors and associated confidence intervals for the CS g-formula and IPW estimators, and to assess the performance of the CS method under varying exposure reliability, unknown, estimated measurement error covariance, near positivity violation, and multiplicative measurement error.

\subsection{G-Formula variance estimation}
\label{sec:gfmla-var}

In the setting of the first simulation study described in the main text, the variance estimators for the oracle, naive, and g-formula CS estimators were evaluated for the single point $\eta(1)$ on the dose-response curve. Web Table \ref{tab:gfmla} reports the empirical bias, empirical standard error (ESE), average estimated standard error (ASE), and the percentage of confidence intervals (CIs) that include the true value. ASE and CI coverage are shown for both the uncorrected (UC) and bias-corrected (BC) variance estimators.

\begin{table}[]
    \centering
    \caption{Detailed results from the first simulation study of the main text. UC: uncorrected empirical sandwich variance estimator, BC: bias-corrected empirical sandwich variance estimator, n: sample size; Bias: 100 times the average bias across simulated data sets for each method; ESE: 100 times the standard deviation of parameter estimates; ASE: 100 times the average of estimated standard errors; Cov: Empirical percent coverage of 95$\%$ confidence intervals for each method.}
    \begin{tabular}{ccrrrrrr}
    \toprule
    \multicolumn{4}{c}{ } & \multicolumn{2}{c}{\textbf{UC}} & \multicolumn{2}{c}{\textbf{BC}} \\
    \cmidrule(l{3pt}r{3pt}){5-6} \cmidrule(l{3pt}r{3pt}){7-8}
    \textbf{n} & \textbf{Method} & \textbf{Bias} & \textbf{ESE} & \textbf{ASE} & \textbf{Cov} & \textbf{ASE} & \textbf{Cov}\\
       \midrule
       400 & Oracle G-Formula & 0.0 & 1.9 & 2.0 & 95.0 & 2.1 & 95.6\\
         & Naive G-Formula & -2.9 & 2.5 & 2.4 & 74.4 & 3.3 & 86.1\\
         & CS G-Formula & 1.7 & 4.0 & 3.4 & 89.8 & 7.5 & 97.6\\
       \addlinespace
       800 & Oracle G-Formula & 0.0 & 1.4 & 1.4 & 95.3 & 1.4 & 95.8\\
         & Naive G-Formula & -2.9 & 1.9 & 1.8 & 59.9 & 2.2 & 70.7\\
         & CS G-Formula & 0.9 & 2.8 & 2.5 & 90.1 & 4.5 & 97.0\\
       \addlinespace
       8000 & Oracle G-Formula & 0.0 & 0.4 & 0.4 & 94.8 & 0.4 & 94.8\\
         & Naive G-Formula & -2.8 & 0.6 & 0.6 & 1.0 & 0.7 & 1.4\\
         & CS G-Formula & 0.2 & 1.0 & 0.9 & 93.2 & 1.0 & 95.3\\
       \bottomrule
                \end{tabular}
    \label{tab:gfmla}
\end{table}

For the oracle estimator, both the UC and BC estimated standard errors were approximately unbiased and the corresponding CIs achieved nominal coverage. The coverages of the CIs corresponding to the naive g-formula estimator were below the nominal level and worsened with increasing sample size, primarily due to bias of this estimator. For the CS g-formula estimator, the UC estimated standard errors tended to underestimate the true standard error, leading to slight undercoverage of the corresponding CIs, with coverage improving as the sample size increased. The BC estimated standard errors for the CS g-formula estimator on the other hand tended to overestimate the true standard errors, leading to slight overcoverage of the corresponding CIs, which also improved with larger $n$.

\subsection{IPW variance estimation}
\label{sec:ipw-var}

The empirical bias, ESE, ASE, and CI coverage for the oracle, naive, and CS IPW estimators in the setting of the second simulation study are presented in Web Table~\ref{tab:ipw}. For all three estimators, both the UC and BC standard error estimators were approximately unbiased for the true standard errors. The CIs corresponding to the naive IPW estimator had low coverage due the bias of this estimator. The CIs corresponding to the oracle and CS IPW estimators achieved nominal coverage.

\begin{table}[]
    \caption{Detailed results from the second simulation study of the main text. UC, BC, n, Bias, ASE, ESE, and Cov defined as in Web Table \ref{tab:gfmla}.}
    \begin{center}
\begin{tabular}{cccrrrrcc}
\toprule
\multicolumn{5}{c}{ } & \multicolumn{2}{c}{\textbf{UC}} & \multicolumn{2}{c}{\textbf{BC}} \\
\cmidrule(l{3pt}r{3pt}){6-7} \cmidrule(l{3pt}r{3pt}){8-9}
\textbf{n} & \textbf{Method} & \textbf{Parameter} & \textbf{Bias} & \textbf{ESE} & \textbf{ASE} & \textbf{Cov} & \textbf{ASE} & \textbf{Cov}\\
 \midrule
 400 & Oracle IPW & $\gamma_0$ & 0.2 & 5.9 & 5.6 & 94.2 & 5.6 & 94.2\\
  &  & $\gamma_1$ & -0.1 & 5.8 & 5.3 & 93.5 & 5.4 & 93.5\\
  &  & $\gamma_2$ & -0.4 & 5.4 & 5.3 & 94.8 & 5.4 & 94.9\\
  \addlinespace
  & Naive IPW & $\gamma_0$ & 0.3 & 6.4 & 6.3 & 95.2 & 6.3 & 95.2\\
  &  & $\gamma_1$ & -16.8 & 5.8 & 5.5 & 14.5 & 5.6 & 14.7\\
  &  & $\gamma_2$ & -16.7 & 5.6 & 5.5 & 15.0 & 5.5 & 15.1\\
  \addlinespace
  & CS IPW & $\gamma_0$ & 0.3 & 6.7 & 6.6 & 95.0 & 6.6 & 95.0\\
  &  & $\gamma_1$ & 0.5 & 7.4 & 7.0 & 93.1 & 7.1 & 93.3\\
  &  & $\gamma_2$ & 0.5 & 7.3 & 7.0 & 94.4 & 7.1 & 94.5\\
 \addlinespace
 800 & Oracle IPW & $\gamma_0$ & 0.0 & 4.3 & 4.0 & 93.9 & 4.0 & 93.9\\
  &  & $\gamma_1$ & -0.2 & 3.9 & 3.8 & 95.0 & 3.8 & 95.0\\
  &  & $\gamma_2$ & 0.1 & 4.0 & 3.8 & 94.0 & 3.8 & 94.1\\
  \addlinespace
  & Naive IPW & $\gamma_0$ & 0.0 & 4.8 & 4.5 & 94.3 & 4.5 & 94.3\\
  &  & $\gamma_1$ & -16.9 & 4.0 & 3.9 & 1.2 & 4.0 & 1.2\\
  &  & $\gamma_2$ & -16.8 & 4.2 & 4.0 & 1.6 & 4.0 & 1.6\\
  \addlinespace
  & CS IPW & $\gamma_0$ & 0.0 & 5.1 & 4.7 & 93.6 & 4.7 & 93.6\\
  &  & $\gamma_1$ & 0.1 & 5.2 & 5.0 & 95.5 & 5.0 & 95.5\\
  &  & $\gamma_2$ & -0.1 & 5.5 & 5.0 & 94.5 & 5.0 & 94.9\\
 \addlinespace
 8000 & Oracle IPW & $\gamma_0$ & 0.0 & 1.5 & 1.3 & 95.9 & 1.3 & 95.9\\
  &  & $\gamma_1$ & 0.0 & 1.4 & 1.3 & 95.3 & 1.3 & 95.3\\
  &  & $\gamma_2$ & 0.0 & 1.3 & 1.2 & 95.0 & 1.2 & 95.0\\
  \addlinespace
  & Naive IPW & $\gamma_0$ & 0.0 & 1.5 & 1.5 & 95.0 & 1.5 & 95.0\\
  &  & $\gamma_1$ & -16.7 & 1.5 & 1.3 & 0.0 & 1.3 & 0.0\\
  &  & $\gamma_2$ & -16.7 & 1.3 & 1.3 & 0.0 & 1.3 & 0.0\\
  \addlinespace
  & CS IPW & $\gamma_0$ & -0.1 & 1.7 & 1.6 & 95.4 & 1.6 & 95.4\\
  &  & $\gamma_1$ & 0.0 & 2.0 & 1.7 & 94.5 & 1.7 & 94.5\\
  &  & $\gamma_2$ & 0.0 & 1.8 & 1.7 & 94.7 & 1.7 & 94.8\\
 \bottomrule
\end{tabular}
\label{tab:ipw}
\end{center}
\end{table}

\subsection{Varying exposure reliability}
\label{sec:exp-rel}

\begin{figure}
\centering
\includegraphics[width=6in]{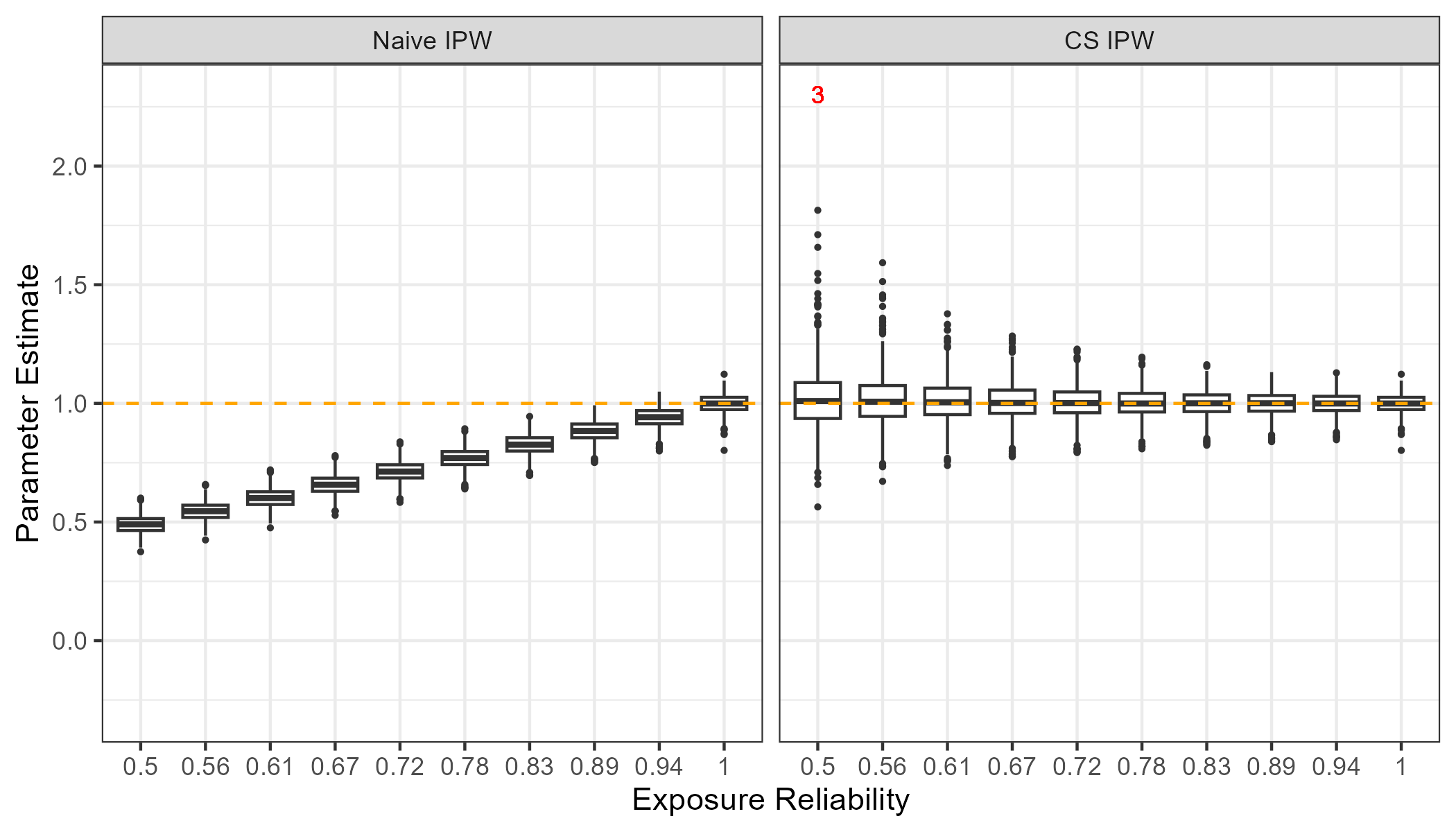}
\caption{Empirical distribution of estimated $\gamma_1$ versus exposure reliability $\Var(A_1)/\Var(A_1^*)=\Var(A_2)/\Var(A_2^*)$, using naive and corrected score (CS) IPW estimators. For the CS IPW estimator and reliability 0.5, three parameter estimates are larger than 2.}
\label{fig:rel1}
\end{figure}

\begin{figure}
\centering
\includegraphics[width=6in]{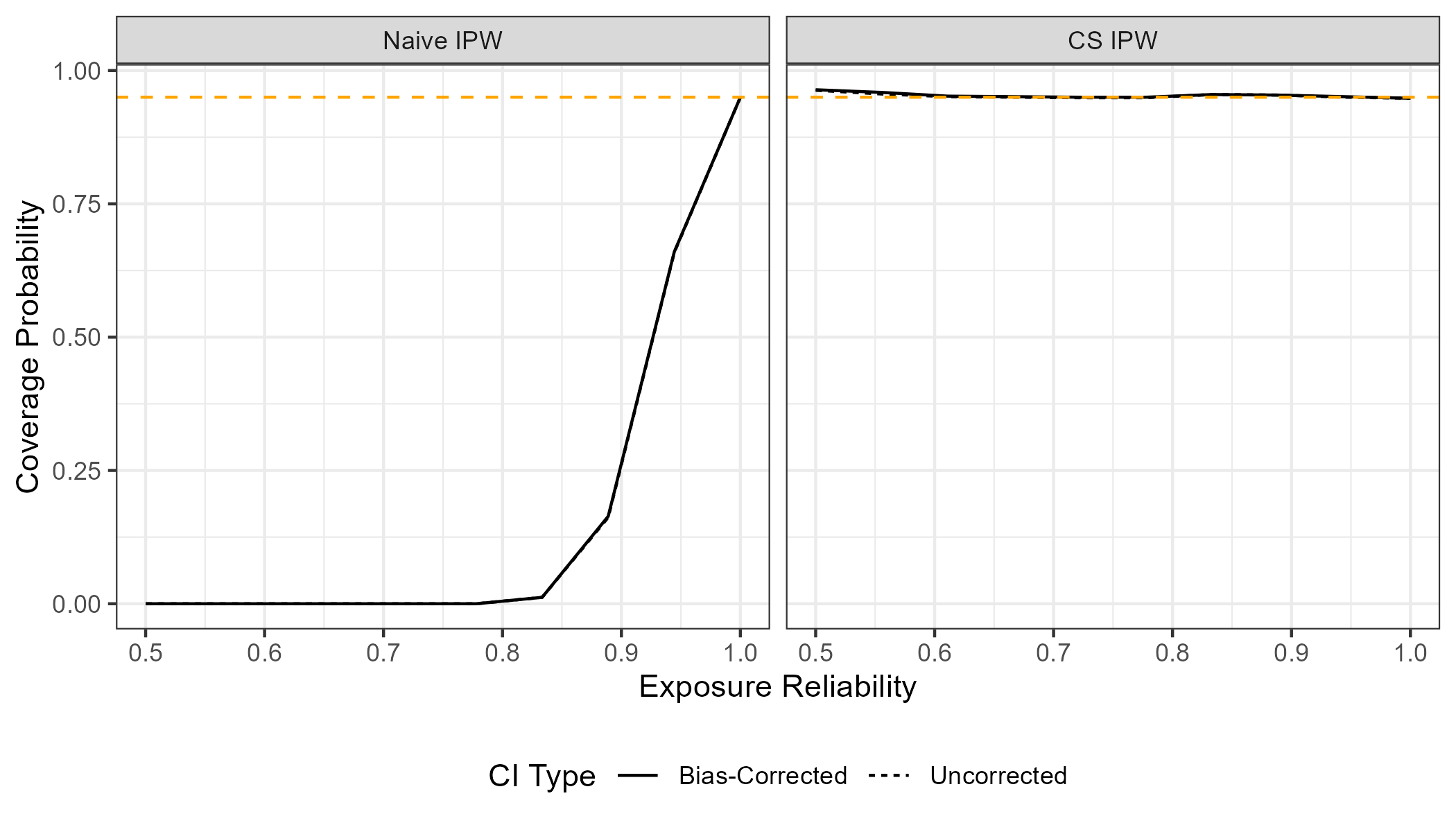}
\caption{Empirical coverage probabilities of 95\% confidence intervals for $\gamma_1$ versus exposure reliability $\Var(A_1)/\Var(A_1^*)=\Var(A_2)/\Var(A_2^*)$, corresponding to naive and corrected score (CS) IPW estimators.}
\label{fig:rel2}
\end{figure}

To evaluate the performance of the proposed CS methods as the exposure measurement error grows, the second simulation study from the main text was repeated over a sequence of exposure reliabilities $\Var(A_1)/\Var(A_1^*) = \Var(A_2)/\Var(A_2^*) \in [0.5,1]$ for sample size $n=800$. Web Figure \ref{fig:rel1} shows the empirical distribution of the naive and CS IPW estimators of $\gamma_1$ versus exposure reliability. Web Figure \ref{fig:rel2} shows the coverage probabilities of the corresponding CIs, both bias-corrected (BC) and uncorrected (UC), versus reliability. 

As the exposure becomes less reliable, the bias of the naive IPW estimator increases and the coverage probabilities of the corresponding CIs drop significantly below the nominal level. The CS IPW estimator, on the other hand, remains approximately unbiased across the sequence of reliabilities, but has greater variance given a less reliable exposure. For all reliabilities in $[0.5, 1]$, the CIs for the CS IPW estimator have approximately nominal coverage.

\subsection{Estimated measurement error covariance}
\label{sec:est-var}

\begin{figure}
\centering
\includegraphics[width=6in]{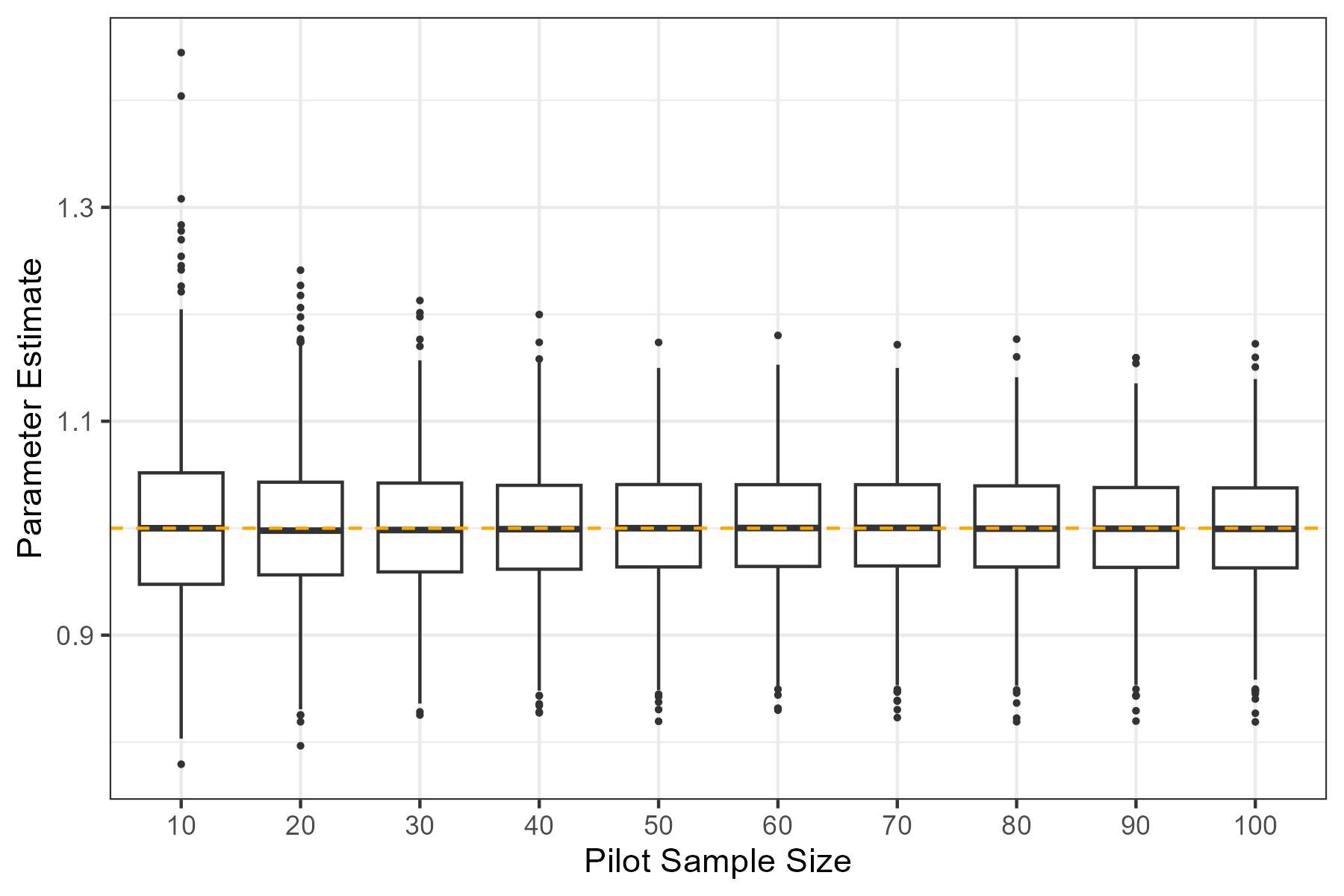}
\caption{Empirical distribution of estimated $\gamma_1$ versus pilot study sample size $n_p$, using the corrected score (CS) IPW estimator with estimated measurement error covariance.}
\label{fig:estvar1}
\end{figure}

\begin{figure}
\centering
\includegraphics[width=6in]{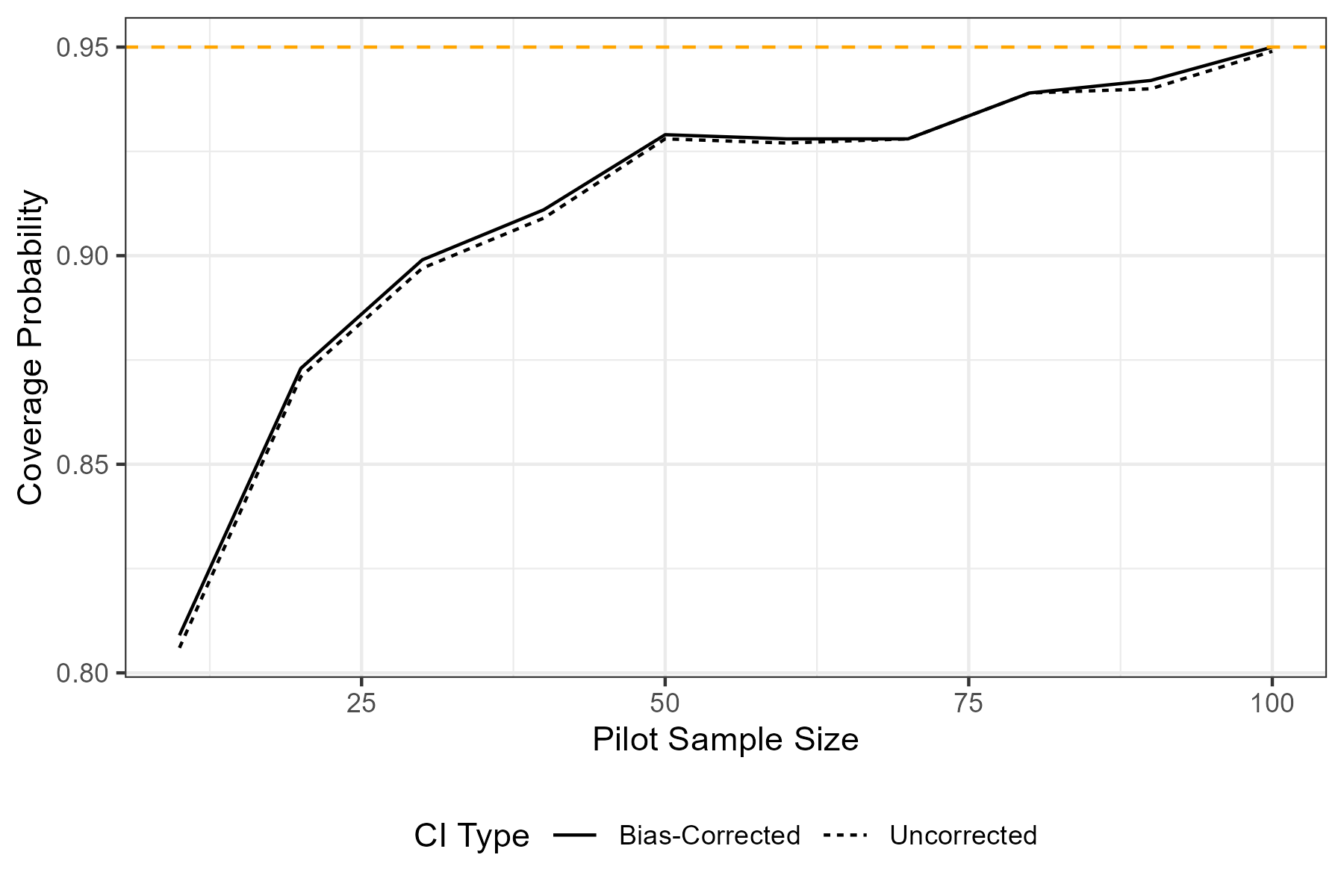}
\caption{Empirical coverage probabilities of 95\% confidence intervals for $\gamma_1$ versus pilot study sample size $n_p$, using the corrected score (CS) IPW estimator with estimated measurement error covariance.}
\label{fig:estvar2}
\end{figure}

The performance of the proposed CS methods was evaluated in the case where an estimated measurement error covariance was used. To study this scenario, the second simulation study from the main text was repeated with an estimated $\widehat{\bSigma}_{me}$ in place of $\bSigma_{me}$. It was assumed that $k=5$ replicate measures of $\bAs$ were observed for $n_p$ individuals in a separate pilot study. Then $\widehat{\bSigma}_{me}$ was computed using Equation \eqref{eq:estvar} from Web Appendix \ref{s:estvar}. To vary the precision of $\widehat{\bSigma}_{me}$, different pilot study sample sizes $n_p$ were considered. Web Figures \ref{fig:estvar1} and \ref{fig:estvar2} show the empirical distribution of CS IPW estimates of $\gamma_1$ and coverage probabilities of corresponding 95\% CIs versus $n_p$, for sample size $n = 800$.

Regardless of pilot study sample size $n_p$, the CS IPW estimator of $\gamma_1$ is approximately unbiased. However, for smaller values of $n_p$, the coverage probabilities of the corresponding CIs drop below the nominal level. This suggests that, if a pilot study of small sample size is used to estimate $\bSigma_{me}$, uncertainty in $\bSigmahat_{me}$ should be considered in the calculation of CIs for $\bgam$, as discussed in Section 3.7 of the main text.

\subsection{Near positivity violation}
\label{sec:pos-vi}

\begin{figure}
\centering
\includegraphics[width=6in]{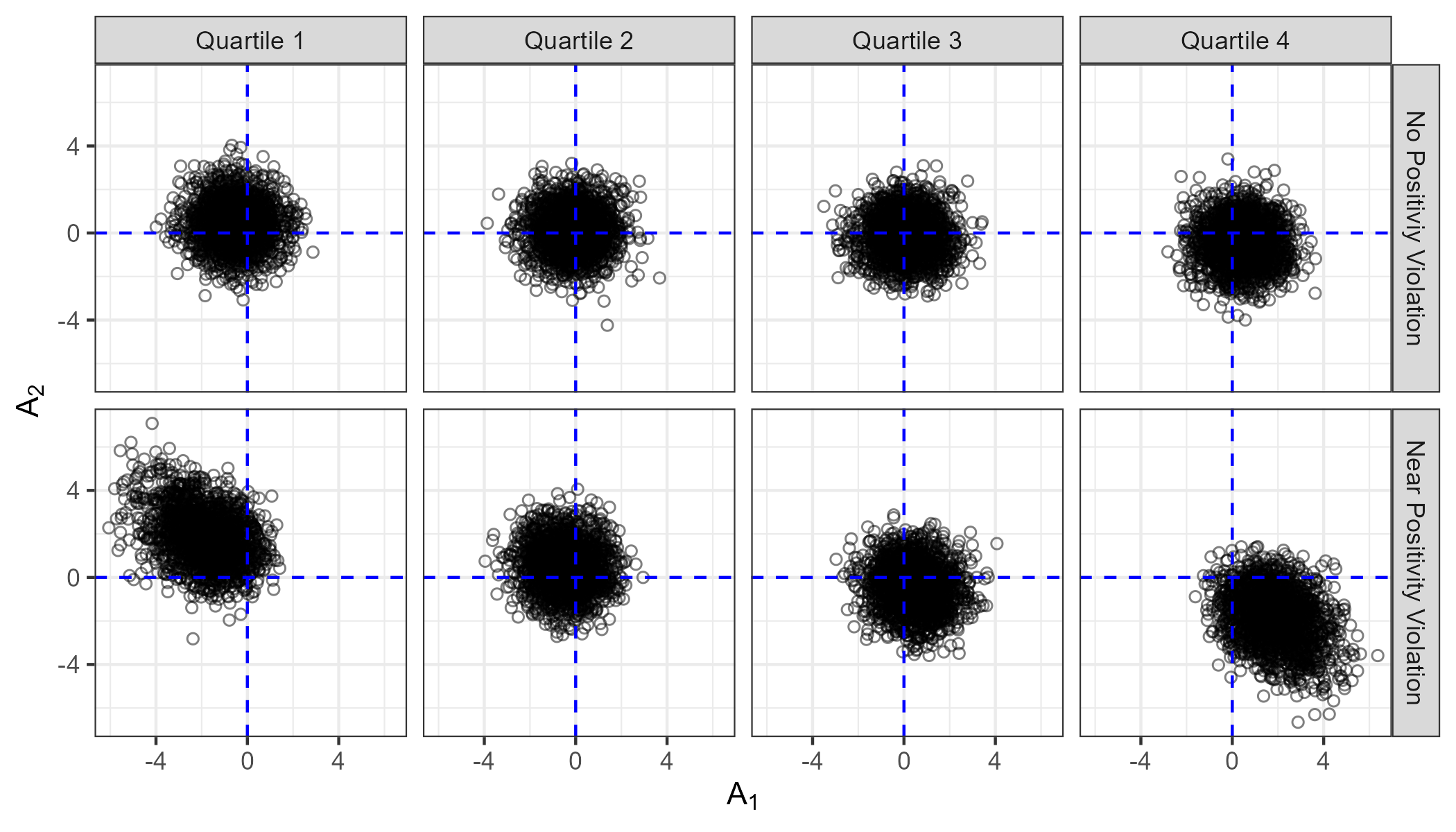}
\caption{True exposure values $(A_1,A_2)$ by quartile of $L$ from one simulated data set of sample size $n=8000$, with and without a near positivity violation.}
\label{fig:pos_setup}
\end{figure}

To evaluate the proposed CS IPW method under a near positivity violation, the general structure of the second simulation study from Section 4 of the main text was replicated almost exactly. A near positivity violation was created by changing the conditional mean of $\bA | L$ from $(L^2, -L^2)$ to $(4L^2, -4L^2)$. Under this new data generating process, which is illustrated and compared to the original data generating process in Web Figure \ref{fig:pos_setup}, values of $(A_1, A_2)$ are almost entirely in the upper left quadrant for small values of $L$ and in the lower right quadrant for large values of $L$. This is termed a near positivity violation because, while strictly speaking the support of $\bA|L$ does not vary with $L$, the conditional density of $\bA|L$ is extremely small for certain values of $L$.

The results of the simulation study are presented in Web Table \ref{tab:pos}. Under this near positivity violation, all three IPW estimators (oracle, naive, and CS) perform poorly, with large bias and low CI coverage that do not resolve with increasing sample size. These results highlight the importance of assessing the positivity assumption, regardless of whether an exposure is mismeasured.

\begin{table}[]
    \caption{Results from the simulation study with a near positivity violation. UC, BC, n, Bias, ASE, ESE, and Cov defined as in Web Table \ref{tab:gfmla}.}
    \begin{center}
\begin{tabular}{cccrrrrcc}
\toprule
\multicolumn{5}{c}{ } & \multicolumn{2}{c}{\textbf{UC}} & \multicolumn{2}{c}{\textbf{BC}} \\
\cmidrule(l{3pt}r{3pt}){6-7} \cmidrule(l{3pt}r{3pt}){8-9}
\textbf{n} & \textbf{Method} & \textbf{Parameter} & \textbf{Bias} & \textbf{ESE} & \textbf{ASE} & \textbf{Cov} & \textbf{ASE} & \textbf{Cov}\\
 \midrule
 800 & Oracle IPW & $\gamma_0$ & 0.2 & 14.6 & 9.1 & 87.8 & 9.4 & 88.7\\
  & & $\gamma_1$ & 4.7 & 10.6 & 6.6 & 72.6 & 7.5 & 77.2\\
  &  & $\gamma_2$ & -5.4 & 9.9 & 6.6 & 72.1 & 7.4 & 77.0\\
 \addlinespace
  & Naive IPW & $\gamma_0$ & 0.8 & 15.3 & 10.2 & 88.2 & 10.5 & 89.2\\
  &  & $\gamma_1$ & -12.6 & 11.0 & 6.8 & 47.6 & 7.8 & 52.3\\
  &  & $\gamma_2$ & -21.2 & 9.9 & 6.8 & 20.1 & 7.7 & 23.0\\
 \addlinespace
  & CS IPW & $\gamma_0$ & 1.0 & 41.4 & 44.4 & 87.9 & 56.3 & 89.7\\
  &  & $\gamma_1$ & 10.0 & 76.1 & 77.5 & 74.8 & 103.7 & 79.2\\
  &  & $\gamma_2$ & -4.8 & 88.2 & 86.7 & 82.8 & 136.2 & 84.9\\
 \addlinespace
 8000 & Oracle IPW & $\gamma_0$ & -0.8 & 10.8 & 6.4 & 90.4 & 6.6 & 91.0\\
  &  & $\gamma_1$ & 4.2 & 8.2 & 4.5 & 65.6 & 5.2 & 69.9\\
  &  & $\gamma_2$ & -3.6 & 8.2 & 4.7 & 64.3 & 5.4 & 68.7\\
 \addlinespace
  & Naive IPW & $\gamma_0$ & -0.4 & 10.7 & 6.8 & 89.1 & 7.0 & 89.5\\
  &  & $\gamma_1$ & -13.0 & 7.9 & 4.6 & 17.9 & 5.1 & 21.6\\
  &  & $\gamma_2$ & -19.9 & 8.0 & 4.6 & 10.2 & 5.1 & 12.2\\
 \addlinespace
  & CS IPW & $\gamma_0$ & -2.3 & 46.0 & 44.9 & 89.9 & 74.3 & 90.1\\
  &  & $\gamma_1$ & 4.4 & 66.4 & 61.9 & 68.0 & 106.0 & 71.5\\
  &  & $\gamma_2$ & -4.3 & 33.0 & 31.0 & 76.6 & 49.8 & 78.8\\
 \bottomrule
\end{tabular}
\label{tab:pos}
\end{center}
\end{table}

\subsection{Multiplicative measurement error}
\label{sec:multiplicative}

\begin{figure}
\centering
\includegraphics[width=6in]{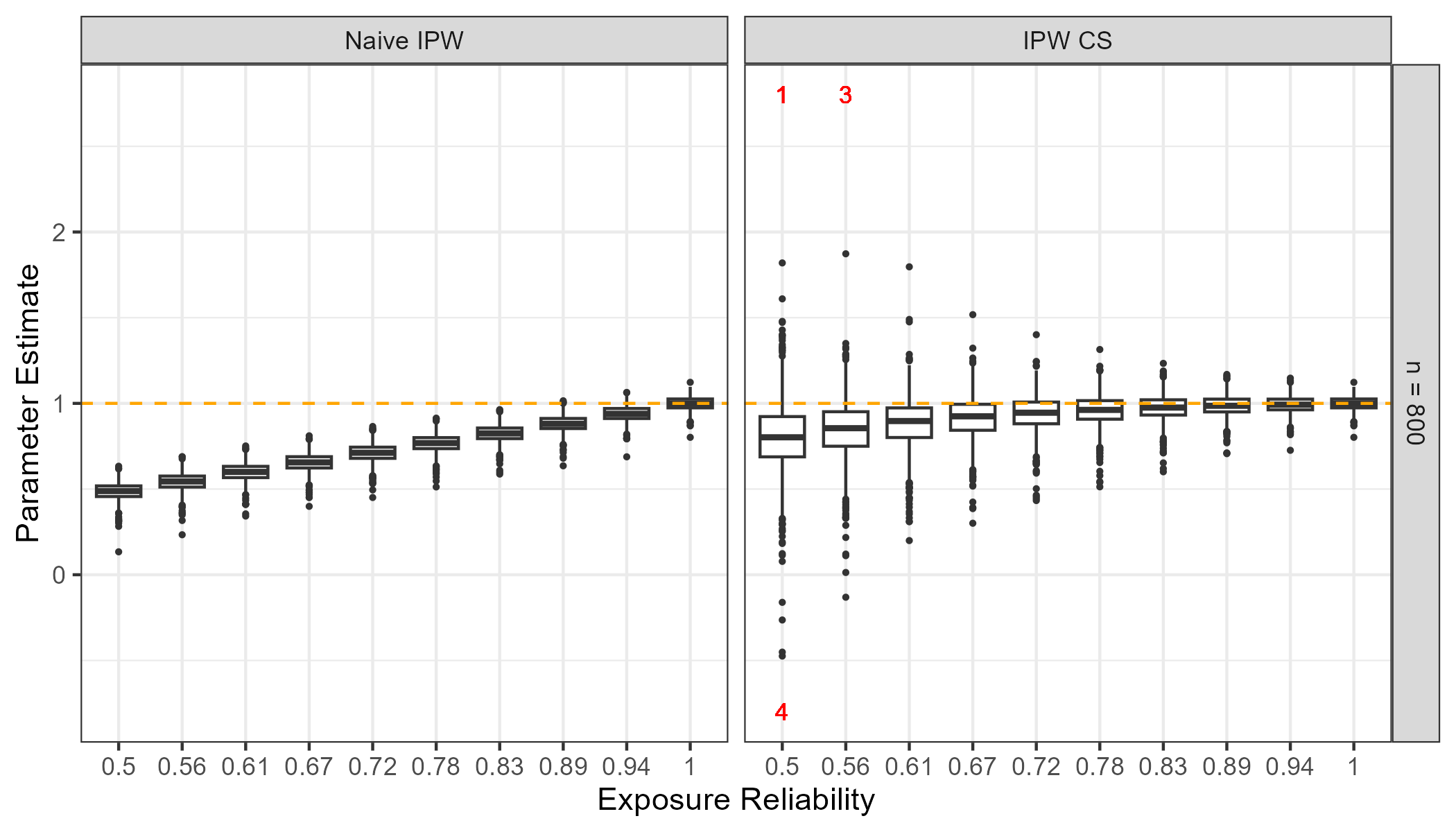}
\caption{Empirical distribution of estimated $\gamma_1$ versus exposure reliability $\Var(A_1)/\Var(A_1^*)=\Var(A_2)/\Var(A_2^*)$, using naive and corrected score (CS) IPW estimators and for a multiplicative measurement error data generating process. For the CS IPW estimator, the number values beyond the ranges of the plot are shown in red on the tails of the boxplots.}
\label{fig:multiplicative1}
\end{figure}

\begin{figure}
\centering
\includegraphics[width=6in]{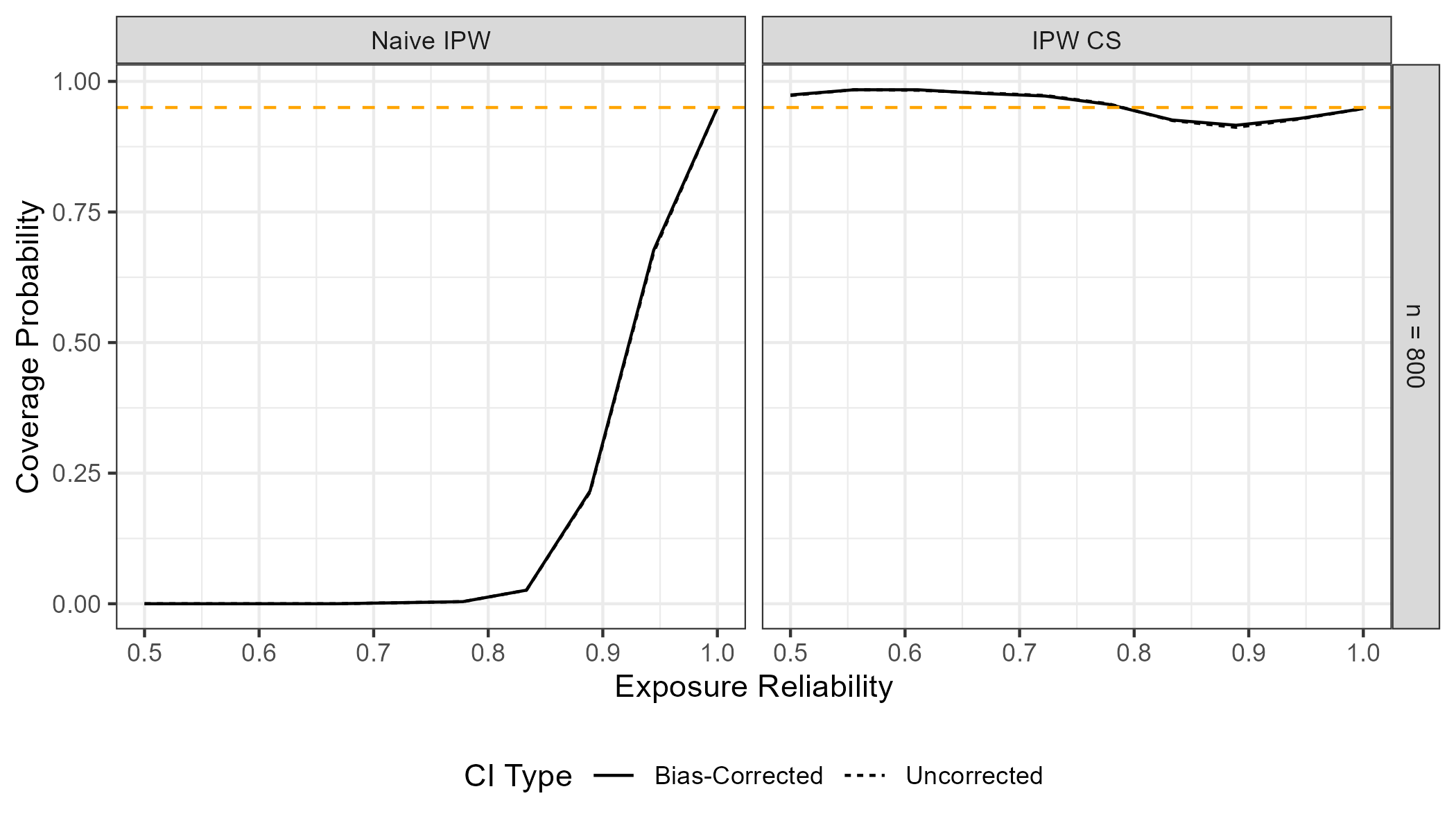}
\caption{Empirical coverage probabilities of 95\% confidence intervals for $\gamma_1$ versus exposure reliability $\Var(A_1)/\Var(A_1^*)=\Var(A_2)/\Var(A_2^*)$, using naive and corrected score (CS) IPW estimators for a multiplicative measurement error data generating process.}
\label{fig:multiplicative2}
\end{figure}

The proposed methods were evaluated when treatment measurement error did not follow the assumed classical additive model. In particular, the second simulation study from Section 4 of the main text was replicated, but with multiplicative measurement error simulated for the exposure. The mismeasured exposure $\bAs = (A^*_1, A^*_2)$ was simulated as $(A_1\epsilon_{me_1}, A_2\epsilon_{me_2})$, where $\beps_{me}=(\epsilon_{me_1}, \epsilon_{me_2})$ followed a bivariate normal distribution with mean $(1,1)$ and diagonal covariance matrix with diagonal elements $(\sigma^2_{me},\sigma^2_{me})$. The measurement error variance $\sigma^2_{me}$ was varied to yield different exposure reliabilities in $[0.5, 1]$. The CS IPW estimator, which incorrectly assumed an additive measurement error model, was implemented using an estimated measurement error covariance (as in Web Appendix \ref{sec:est-var}) with $k=5$ and $n_p=100$. For comparison, the naive IPW estimator was also used. Web Figures \ref{fig:multiplicative1} and \ref{fig:multiplicative2} show the empirical distribution of the naive and CS estimators of $\gamma_1$, and empirical coverage probabilities of the corresponding CIs, versus exposure reliability.

For lower exposure reliabilities, the CS IPW exhibited bias, which is expected since it used a mis-specified measurement error model. However, in terms of both bias and CI coverage probability, the CS IPW method outperformed the naive method across the entire range of exposure reliabilities.

\section{Diagnostics and supplemental analyses for application section}

\subsection{Model diagnostics for the application}

\begin{figure}
\centering
\includegraphics[width=5in]{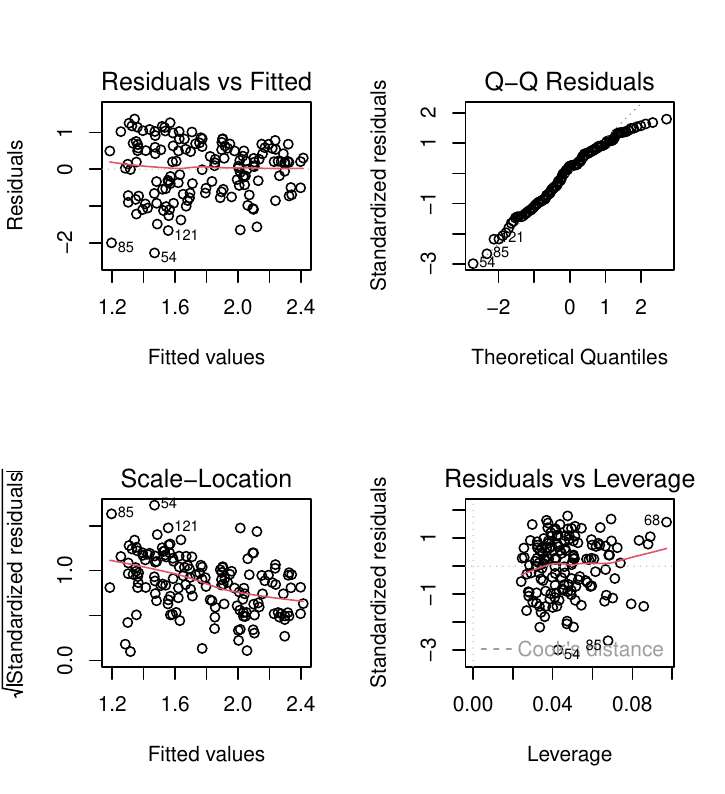}
\caption{Model diagnostics for the ADCP propensity model.}
\label{fig:diag-adcp}
\end{figure}

\begin{figure}
\centering
\includegraphics[width=5in]{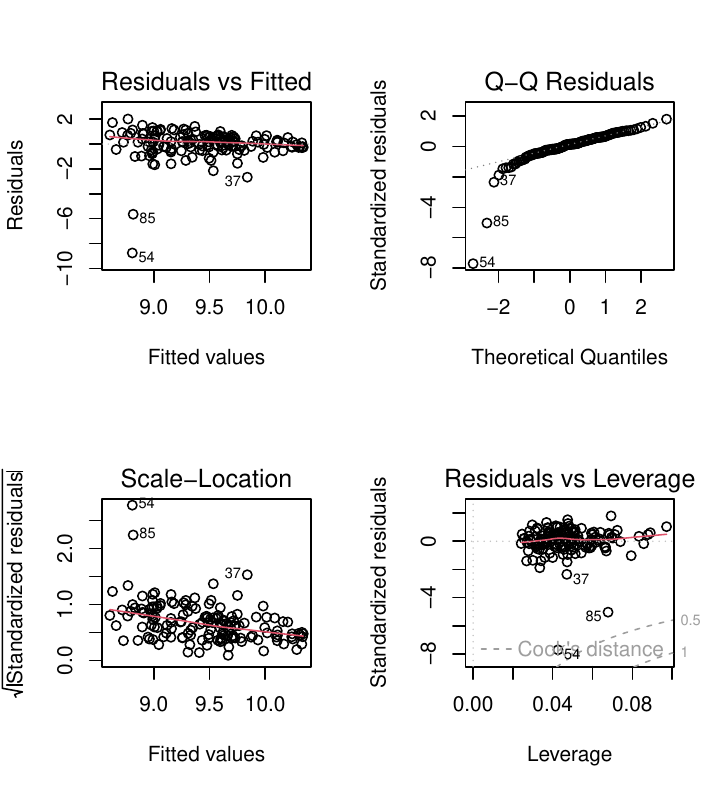}
\caption{Model diagnostics for the R\RNum{2} propensity model.}
\label{fig:diag-rII}
\end{figure}

Standard model diagnostics were used to evaluate both propensity model specifications in Section 5 of the main text. Diagnostics for the ADCP exposure are presented in Web Figure \ref{fig:diag-adcp}, which largely indicate a good model fit. Diagnostics for the R\RNum{2} exposure are presented in Web Figure \ref{fig:diag-rII}. These diagnostics indicate two outliers for which the model had poor predictive performance.

\begin{figure}
\centering
\includegraphics[width=6.5in]{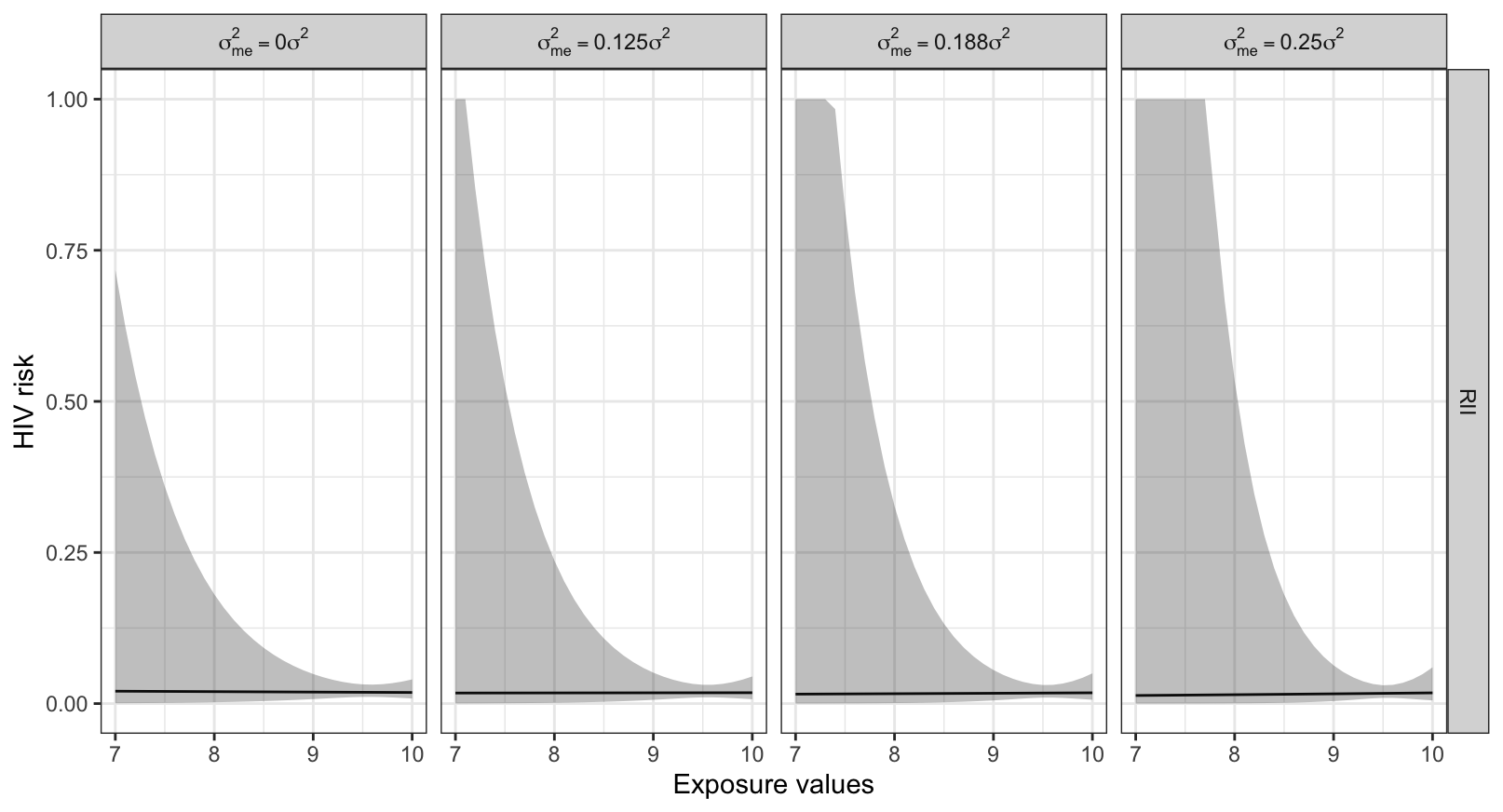}
\caption{Supplemental analysis removing two outliers in the R\RNum{2} models.}
\label{fig:rII2}
\end{figure}

\begin{figure}
\centering
\includegraphics[width=5in]{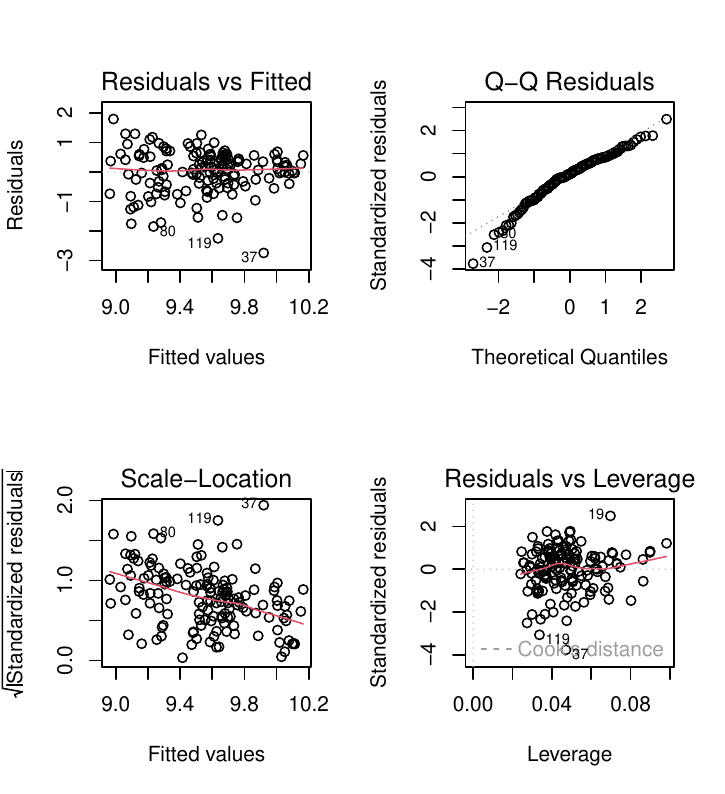}
\caption{Model diagnostics for the R\RNum{2} propensity model in the supplemental analysis with two outliers removed}.
\label{fig:diag-rII2}
\end{figure}

To assess the fit of the outcome models, Chi-square goodness of fit tests were performed, yielding a p-value of 0.56 for the model including ADCP and a p-value of 0.33 for the model including R\RNum{2}. Based on the propensity model diagnostics, a supplemental analysis was performed with two outliers removed for the R\RNum{2} models (Web Figure \ref{fig:rII2}). The diagnostics for this analysis no longer indicated poor propensity model fit (Web Figure \ref{fig:diag-rII2}) and yielded an outcome model Chi-square goodness of fit test p-value of 0.85. However, removing the outliers resulted in wider confidence intervals for lower levels of the exposure than in the analysis presented in Section 5 of the main paper.

\subsection{Application with nonlinear outcome model}

\begin{figure}
\centering
\includegraphics[width=6.5in]{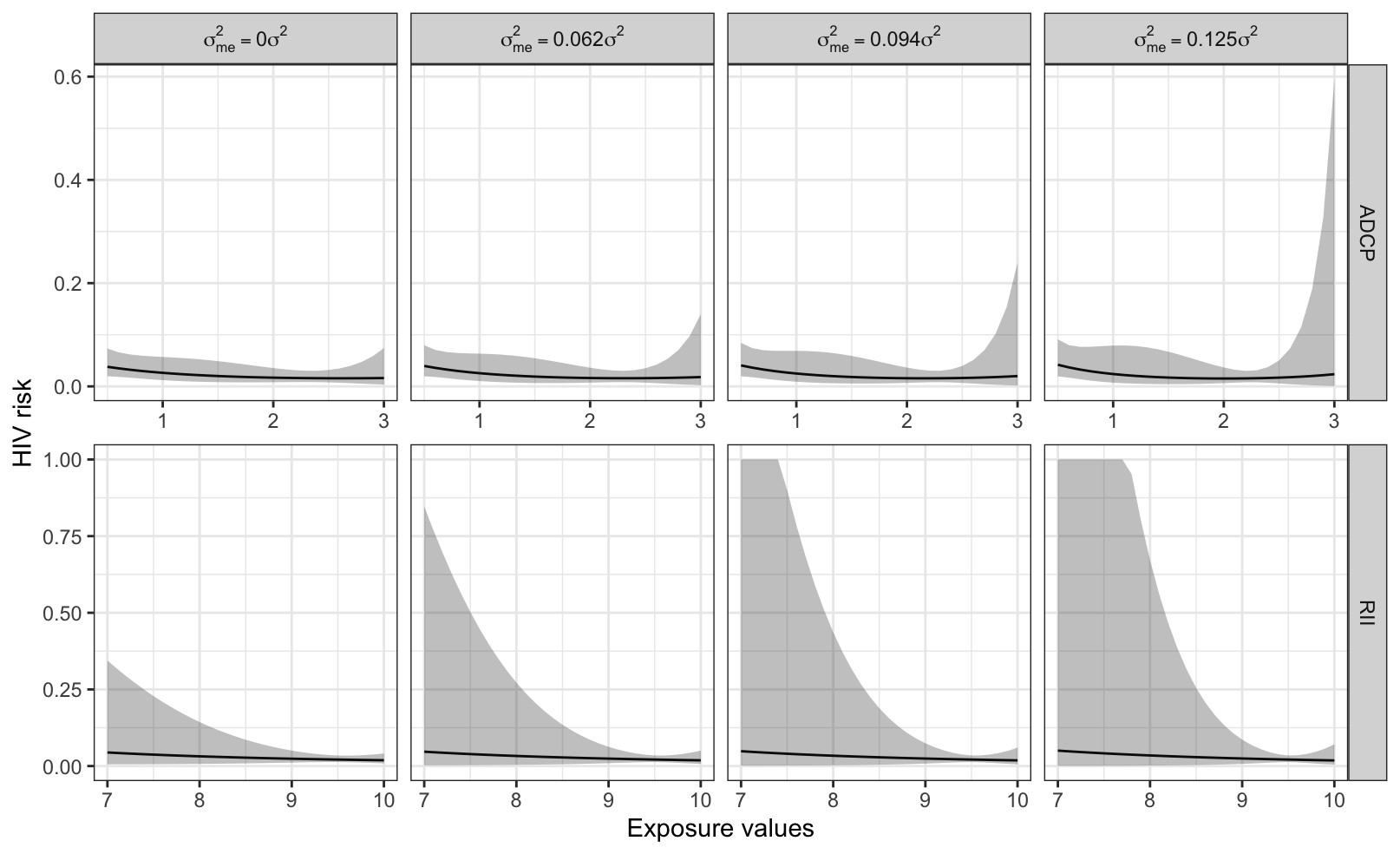}
\caption{Results for the supplemental analyses using the CS g-formula estimator and a quadratic outcome model.}
\label{fig:supp-nonlinear}
\end{figure}

A second supplementary analysis of the HVTN 505 application was performed assuming a nonlinear outcome model. In particular, the proposed g-formula estimator was used specifying the same outcome model form as described for the doubly-robust estimator used in the main text, but with an additional quadratic term for the exposure of interest (ADCP or R\RNum{2}). A slightly smaller range of measurement error variances (equal to 0, 1/16, 3/32, and 1/8 times the observed exposure variance) was considered, noting that large error variances can lead to less stability and wider confidence regions when models are specified with quadratic or higher-order polynomial terms. The results are presented in Web Figure \ref{fig:supp-nonlinear}.

\section{Two-phase sampling}
\label{sec:two-phase}

\subsection{Two-phase sampling method}
\label{sec:two-phase-method}

Many studies (including the HVTN 505 trial) use a two-phase sampling design. Such a design is particularly useful when the outcome is easy to measure but the exposure of interest or some covariates are expensive or difficult to measure. Because each of the proposed methods above belongs to the estimating equation framework, it is straightforward to incorporate previously described methods for causal inference from studies with two-phase sampling. In this section, one such approach is demonstrated using a simulation study. In particular, for this simulation and the application section analysis, a simple inverse probability of sampling weights method is used~\citep{wang2009}.

The method is implemented by weighting each individual's contribution to the estimating equations by the inverse probability of selection for the second-phase of the study (multiplying treatment weights by sampling weights for the CS IPW and CS DR estimators) and restricting the analysis to those selected. This method is well-suited for the subset of the HVTN 505 trial that is the focus of Section 5 of the main paper, particularly because all exposures of interest were measured in the second-phase sub-sample and no exposures were measured in the full sample.

\subsection{Two-phase sampling simulations}
\label{sec:two-phase-sims}

The third simulation study described in Section 4 of the main text was replicated, but under a two-phase sampling design. In particular, a case-cohort design was used where the exposure was measured for a random sub-cohort of controls as well as for every case. This was done for a sample size of $n=2000$ under five scenarios, with sub-cohorts of size $5\%$, $10\%$, $25\%$, $50\%$ and $100\%$. A sub-cohort size of $100\%$ corresponds to the setting of the third simulation study of the main text, where no case-cohort sampling is used. The estimand of interest ($\gamma_1$) was the same as in main text.

\begin{table}
    \centering
    \footnotesize
    \caption{Results from the two-phase sampling simulation study. UC, BC, Bias, ASE, ESE, and Cov defined as in Web Table \ref{tab:gfmla}.}
   \begin{tabular}{cccrrrrr}
\toprule
\multicolumn{4}{c}{ } & \multicolumn{2}{c}{\textbf{UC}} & \multicolumn{2}{c}{\textbf{BC}} \\
\cmidrule(l{3pt}r{3pt}){5-6} \cmidrule(l{3pt}r{3pt}){7-8}
\textbf{Sub-cohort size} & \textbf{Method} & \textbf{Bias} & \textbf{ESE} & \textbf{ASE} & \textbf{Cov} & \textbf{ASE} & \textbf{Cov}\\
 \midrule
 5\% & Oracle DR & 1.5 & 20.7 & 17.1 & 87.0 & 17.3 & 87.5\\
  & Naive DR & -5.2 & 16.7 & 14.1 & 88.6 & 14.3 & 89.0\\
  & CS DR & 1.9 & 27.6 & 23.4 & 88.3 & 23.9 & 89.1\\
 \addlinespace
 10\% & Oracle DR & 0.7 & 15.5 & 13.2 & 89.4 & 13.4 & 89.7\\
  & Naive DR & -5.3 & 12.4 & 10.8 & 88.5 & 10.9 & 88.8\\
  & CS DR & 1.5 & 20.8 & 17.6 & 88.5 & 17.9 & 88.8\\
 \addlinespace
 25\% & Oracle DR & 0.7 & 10.0 & 9.3 & 91.8 & 9.3 & 91.8\\
  & Naive DR & -5.4 & 7.8 & 7.4 & 86.7 & 7.5 & 86.9\\
  & CS DR & 1.1 & 13.2 & 12.2 & 91.4 & 12.2 & 91.6\\
 \addlinespace
 50\% & Oracle DR & 0.4 & 7.7 & 7.2 & 92.6 & 7.3 & 92.6\\
  & Naive DR & -5.7 & 6.2 & 5.8 & 82.3 & 5.8 & 82.4\\
  & CS DR & 0.3 & 10.5 & 9.5 & 90.4 & 9.6 & 90.8\\
 \addlinespace
 100\% & Oracle DR & 0.0 & 6.2 & 6.0 & 94.0 & 6.0 & 94.0\\
  & Naive DR & -5.8 & 4.9 & 4.7 & 76.6 & 4.7 & 76.6\\
  & CS DR & 0.2 & 8.5 & 7.9 & 92.6 & 7.9 & 92.9\\
 \bottomrule
\end{tabular}
    \label{tab:twophase}
\end{table}

The results of 1000 simulations are presented in Web Table~\ref{tab:twophase}. There was some bias and under-coverage when the sub-cohorts were smaller. However, as the sub-cohort size increased, the bias and standard errors decreased, and the confidence interval coverage corresponding to the oracle and CS DR estimators approached 95\%.

\label{lastpage}

\end{document}